\documentstyle[aps,prd]{revtex}

\tightenlines

\draft
\begin{document}

\title{Multiple-Scale Analysis and Renormalization of
Quenched Second Order Phase Transitions}

\author{Sang Pyo Kim,$^{1,2,}$\footnote{E-mail:
spkim@phys.ualberta.ca; sangkim@ks.kunsan.ac.kr}
Supratim Sengupta,$^{1,3}$\footnote{E-mail: sengupta@phys.ualberta.ca}
and F. C. Khanna,$^{1,3,}$\footnote{E-mail:
khanna@phys.ualberta.ca}}

\address{${}^{1}$Theoretical Physics Institute, Department of
Physics, University of Alberta, Edmonton, Alberta, Canada T6G
2J1\\ ${}^{2}$Department of Physics, Kunsan National University,
Kunsan 573-701, Korea \\
${}^{3}$TRIUMF, 4004 Wesbrook Mall, Vancouver, British Columbia,
Canada, V6T 2A3}

\date{\today}

\maketitle

\begin{abstract}
A quenched second order phase transition is modeled by an
effective $\Phi^4$-theory with a time-dependent Hamiltonian
$\hat{H} (t)$, whose symmetry is broken spontaneously in time. The
quantum field evolves out of equilibrium (nonequilibrium) during
the phase transition as the density operator significantly
deviates from $\hat{\rho} (t) = e^{- \beta \hat{H} (t)}/Z_H$. The
recently developed Liouville-von Neumann (LvN) method provides
various quantum states for the phase transition in terms of a
complex solution to the mean-field equation, which is equivalent
to the Gaussian effective potential in the static case and to the
time-dependent Hartree-Fock equation in the nonequilibrium case.
Using the multiple scale perturbation theory (MSPT) we solve
analytically the mean-field equation to the first order of
coupling constant and find the quantum states during the quenched
second order phase transition. We propose a renormalization scheme
during the process of phase transition to regularize the
divergences, which originate from the mode coupling between hard
and hard modes or between the soft and hard modes. The effect of
mode coupling is discussed.
\end{abstract}


\section{Introduction}

The dynamics of nonequilibrium (out of equilibrium) quantum fields
has been a subject of much discussion in the recent years
\cite{gleiser,karra}. Nonequilibrium quantum fields play important
roles in a variety of different scenarios. During a second order
phase transition, the time scale of relaxation of the scalar field
lags behind the time scale of the change of the effective
potential. Consequently the field evolves out of equilibrium as it
tries to relax to the new vacuum state, having a nonvanishing
vacuum expectation value. Such nonequilibrium effects play a
crucial role in topological defect formation both in condensed
matter systems as well as in the early universe
\cite{rivers,calzetta,hu}. In a seminal paper, Kibble first showed
how the correlation length is crucial in determining the initial
density of topological defects \cite{kbl}. These ideas were
applied by Zurek who proposed that it may be possible to
quantitatively test the Kibble mechanism of defect formation in
condensed  matter systems like superfluid ${\rm He}^4$
\cite{zurek}. He argued that, due to the phenomenon of critical
slowing down near the phase transition point, the correlation
length relevant for determining the initial density of the defects
is not the equilibrium correlation length at the Ginzburg
temperature but that at the time when the field dynamics
essentially freezes out. He also found a power law behaviour in
the dependence of the correlation length (and consequently initial
defect density) on the quench rate. There have been various
attempts to experimentally test the Kibble-Zurek prediction of
initial defect density \cite{zurek2}. A comprehensive review of
these issues is given in Ref. \cite{karra}.

The nonequilibrium evolution of a scalar field also is responsible
for the establishment of large-scale correlations leading to
growth of domains \cite{boyan}. This is particularly relevant in
the context of Disoriented Chiral Condensate (DCC) formation,
where only a large domain of DCC can produce the dramatic
fluctuations in the charged to neutral pion ratio \cite{wilczek}.
The field theoretical treatment from nonequilibrium to equilibrium
is particularly important in understanding how a system of quarks
and gluons thermalizes after a heavy-ion collision to form a new
state of matter called quark-gluon plasma. Searches of quark-gluon
plasma are currently underway at RHIC in Brookhaven and LHC in
CERN. In another context, the recent preheating mechanism of the
Universe after inflation might have accompanied the explosive
decay of the inflaton into a large number of soft quanta of the
field(s) \cite{linde,yoshi,tkachev,devega}. This process, leading
to a large population of the low momentum modes, occurs out of
equilibrium. A subsequent interaction between the various modes
leads to the transfer of energy from low to high momentum modes
and the eventual thermalization of the Universe. Another exciting
aspect of the nonequilibrium field dynamics which has recently
received much attention is the issue of decoherence and consequent
appearance of classicality during a nonequilibrium phase
transition \cite{lombardo,kim-lee,habib}.

A feature common to all such (quenched) phase transitions is the
evolution far away from equilibrium and the exponential growing of
the soft (long wavelength) modes, which necessarily leads to the
domain growth. Finite temperature field theory based on
equilibrium or quasi-equilibrium methods does not describe all the
processes of nonequilibrium evolution even though the imaginary
part of the complex effective action yields the decay rate
\cite{weinberg}. To treat properly such nonequilibrium quantum
evolution, Schwinger and Keldysh first introduced the closed-time
formalism \cite{schwinger}. The closed-time formalism or the $1/N$
expansion method have been applied to the nonequilibrium
$\Phi^4$-theory to explain the phenomenon of domain growth
\cite{boyan,cooper}. In a recent paper \cite{kim-lee2}, a
canonical method called the Liouville-von Neumann (LvN) approach
has been developed that unifies both the functional
Schr\"{o}dinger equation for quantum evolution and the quantum LvN
equation for quantum statistics. At the lowest order of coupling
parameter the LvN approach yields the Gaussian approximation for
static quantum fields and leads to mean-field equations for
nonequilibrium quantum fields that can be equivalently obtained by
the time-dependent Hartree-Fock method.

The purpose of this paper to study the renormalization of an
effective theory of $\Phi^4$ field undergoing a quenched second
order phase transition. The $\Phi$ may represent a quantum field
of an effective system interacting with an environment or an order
parameter of the system. In all such cases our model for the
second phase transition is described by the Hamiltonian
\begin{equation}
H(t) = \int \frac{d^3{\bf x}}{(2 \pi)^{3/2}}
\Biggl[\frac{1}{2}\Pi^2 + \frac{1}{2} (\nabla \phi)^2 +
\frac{m^2_{\rm B} (t)}{2} \Phi^2 + \frac{\lambda_{\rm B} }{4!}
\Phi^4 \Biggr], \label{ham}
\end{equation}
where
\begin{eqnarray}
 m^2_{\rm B} (t) = \cases{ m_{B}^2, & $ t < 0$, \cr -
\tilde{m}^2_{\rm B}, & $t
> 0$. \cr} \nonumber
\end{eqnarray}
The transition is modeled through a sudden and instantaneous
change in sign of $m^2_{\rm B}$ at $t=0$; this change destabilizes
the field whose true vacuum was at $\Phi = 0$ before $t < 0$ and
signifies the onset of the spontaneous symmetry breaking
transition. Before the phase transition the quantum field is in a
static state of either vacuum state or thermal equilibrium and can
be accurately described by the Gaussian effective potential (GEP)
\cite{chang,stevenson}. But the onset of the quench drives the
field out of equilibrium, so the finite temperature field theory
or GEP may not be directly applied; instead the LvN or the
Hartree-Fock method provides the correct evolution of the field
even after the phase transition. In this paper we shall employ the
LvN method, as this method exactly recovers the GEP result before
the phase transition and the time-dependent Hartree-Fock result
after the phase transition.

The Hartree-Fock method has been popular and useful in studying
the nonequilibrium evolution of the field
\cite{boyan,cooper,ringwald}. Even though the renormalization is
well understood even for nonequilibrium fields in Refs.
\cite{boyan,cooper,ringwald}, a systematic and explicit
renormalization scheme has not been properly addressed for the
phase transitions. In this paper we propose such a renormalization
scheme for phase transitions by applying the multiple scale
perturbation theory (MSPT) method \cite{bender,bender2} to the LvN
method and thus obtain the renormalized parameters after phase
transitions. The MSPT method introduces the multiple time scales
determined by the nonlinear coupling constant of the system, say
$\lambda$, and leads to the solution with the correct renormalized
frequency even at order $\lambda \hbar$. In field theoretical
terms the MSPT solution even at the lowest order is equivalent to
summing selected diagrams to all orders and is an accurate
approximation for a sufficiently long time, $1/(\lambda \hbar)$.
The MSPT method turns out particularly useful within the LvN
method since the procedure of renormalization in the mean-field or
the Hartree-Fock equation amounts to obtaining the renormalized
frequencies of the hard and soft modes according to the MSPT
method to zeroth order. The calculation is carried out within the
Gaussian approximation and, therefore, does not address all the
issues pertaining to the non-Gaussian effect of nonlinear mode
mixing. It nevertheless provides an analytical (albeit,
perturbative) method of obtaining expressions for the renormalized
parameters during a spontaneous symmetry breaking phase
transition.

The paper is organized as follows. In Sec. II, we show how the LvN
method can be used to obtain the renormalized Gaussian effective
potential. We also obtain the appropriate form of the renormalized
mass and coupling constant using the GEP. In Sec. III, we study a
simple quantum-mechanical model of two coupled quartic oscillators
as a precursor to studying the nonequilibrium evolution of the
quantum fields during a quenched second order phase transition. We
apply the MSPT method to obtain the correct frequency for each
mode. Although the issue of renormalization is not relevant for
this quantum mechanical model, it nevertheless provides some
insight into the nonlinear effect of coupling between soft and
hard modes during the quenched second order phase transition. In
Sec. IV, we deal with the renormalization of mass and coupling
parameters during the nonequilibrium evolution of the quantum
fields without phase transitions. In Sec. V, by applying the MSPT
we obtain the renormalized solution to field equations after the
second order phase transition. A summary and discussion of our
results is given in Sec. V.

\section{GEP from the LvN Method}

The essential idea of the LvN method \cite{kim-lee2} is to
simultaneously solve the functional Schr\"{o}dinger equation
\begin{equation}
i \hbar \frac{\partial}{\partial t} \vert \Psi (t) \rangle =
\hat{H} (t) \vert \Psi(t) \rangle, \label{sch eq}
\end{equation}
and the quantum LvN equation
\begin{equation}
i \hbar \frac{\partial}{\partial t} \hat{\cal O} (t) + [\hat{\cal O} (t),
\hat{H} (t)] = 0. \label{ln eq}
\end{equation}
The LvN method is based on the observation by Lewis and Riesenfeld
\cite{lewis} that the eigenstates of $\hat{\cal O} (t)$ are the
exact quantum states of time-dependent Schr\"{o}dinger equation
(\ref{sch eq}) up to time-dependent phase factors. For a
time-independent system the Hamiltonian itself satisfies the LvN
equation (\ref{ln eq}) and its energy eigenstates are exact
quantum states as expected. In terms of appropriately selected
operators satisfying Eq. (\ref{ln eq}) one may find the density
operator and the Hilbert space of exact quantum states. In this
sense the LvN method unifies quantum statistical mechanics with
quantum mechanics. The LvN method treats the time-dependent,
nonequilibrium, system exactly in the same way as the
time-independent, equilibrium, one. Also the LvN method can be
applied to nonequilibrium fermion systems with a minimal
modification \cite{kim-khanna}.

To compare the LvN method with the well-known GEP
\cite{stevenson}, a nonperturbative method, we consider a field
theoretic model to which both methods can apply. Such a model is
provided by the $\Phi^4$-theory with the Hamiltonian
\begin{equation}
H(t) = \int \frac{d^3{\bf x}}{(2 \pi)^{3/2}}
\Biggl[\frac{1}{2}\Pi^2 + \frac{1}{2} (\nabla \Phi)^2 +
\frac{m^2_{\rm B}}{2} \Phi^2 + \frac{\lambda_{\rm B}}{4!} \Phi^4
\Biggr]. \label{free ham}
\end{equation}
In the GEP method the effective potential of a classical
background field $\phi_c$ includes corrections from quantum
fluctuations. The field is assumed to fluctuate in a Gaussian
vacuum state with the expectation value
\begin{equation}
\langle 0 \vert \hat{\Phi} \vert 0 \rangle_{\rm G} = \phi_c, \quad
\langle 0 \vert \hat{\Pi} \vert 0  \rangle_{\rm G} = 0,
\end{equation}
where the field is divided into the classical background field and
fluctuations
\begin{equation}
\Phi (t, {\bf x})  = \phi_c + \Phi_q (t, {\bf x}).
\end{equation}
To express Eq. (\ref{free ham}) as an infinite sum of coupled
quartic oscillators, we redefine the Fourier modes of fluctuation
as
\begin{equation}
\phi_{\bf k}^{(+)} (t) = \frac{1}{2} [\phi_{\bf k} (t) + \phi_{-
{\bf k}} (t)], \quad \phi_{\bf k}^{(-)} (t) =
\frac{i}{2}[\phi_{\bf k} (t) - \phi_{- {\bf k}} (t)], \label{mode}
\end{equation}
where
\begin{equation}
\Phi_q (t, {\bf x}) = \int \frac{d^3 {\bf k}}{(2 \pi)^{3/2}}
\phi_{\bf k} (t) e^{i {\bf k} \cdot {\bf x}}.
\end{equation}
Then the mode-decomposed Hamiltonian is given by
\begin{equation}
H = \sum_{\alpha} \frac{1}{2}\pi_{\alpha}^2 + \frac{1}{2}
\bar{\omega}_{\alpha}^2 \phi_{\alpha}^2 + \frac{\lambda_{\rm B}}{4!}
\Biggl[\sum_{\alpha} \phi_{\alpha}^4 + 3 \sum_{\alpha \neq \beta}
\phi_{\alpha}^2 \phi_{\beta}^2 \Biggr] + \frac{m_{\rm B}^2}{2} \phi_c^2
 + \frac{\lambda_{\rm B}}{4!} \phi_c^4
\label{ham-eff}
\end{equation}
where $\alpha = \{ (\pm), {\bf k}\}$ and
\begin{equation}
\bar{\omega}_{\alpha}^2 = m_{\rm B}^2 + {\bf k}^2 +
\frac{\lambda_{\rm B}}{2} \phi_c^2.
\end{equation}
Here the odd power terms of $\phi_{\alpha}$ are neglected since we
are interested in symmetric quantum states. Following Kim and Lee
\cite{kim-lee2}, we introduce the operators
\begin{eqnarray}
\hat{A}^{\dagger}_{\alpha} (t) = - \frac{i}{\sqrt{\hbar}}
[\varphi_{\alpha} (t) \hat{\pi}_{\alpha} - \dot{\varphi}_{\alpha}
(t) \hat{\phi}_{\alpha} ], \quad \hat{A}_{\alpha} (t) =
\frac{i}{\sqrt{\hbar}} [\varphi^*_{\alpha} (t) \hat{\pi}_{\alpha}
- \dot{\varphi}_{\alpha}^* (t) \hat{\phi}_{\alpha} ],  \label{ln
op}
\end{eqnarray}
with overdots denoting time derivative, and express the
Hamiltonian (\ref{ham-eff}) as normal-ordered quadratic and
quartic parts in $\hat{A}$ and $\hat{A}^{\dagger}$:
\begin{equation}
\hat{H} = \hat{H}_{\rm G} +  \hat{H}_{\rm P},
\end{equation}
where
\begin{eqnarray}
\hat{H}_{\rm  G} &=& \frac{\hbar}{2} \sum_{\alpha} [ (\dot{\varphi
}^{2}_{\alpha} + \bar{\omega}^2_{\alpha} {\varphi}^{2})
\hat{A}^{2}_{\alpha} + (\dot{\varphi}^*_{\alpha}
\dot{\varphi}_{\alpha} + \bar{\omega}^2_{\alpha}
{\varphi}^*_{\alpha} {\varphi}_{\alpha} ) (2
\hat{A}^{\dagger}_{\alpha} \hat{A}_{\alpha} + 1) +
(\dot{\varphi}^{*2}_{\alpha} + \bar{\omega}^2_{\alpha}
{\varphi}^{*2}_{\alpha} )\hat{A}^{\dagger 2}_{\alpha} ]
\nonumber\\ && + \frac{\lambda_{\rm B} \hbar^2}{4} \sum_{{\alpha},
{\beta}} ({\varphi}^*_{\beta} {\varphi}_{\beta})
[{\varphi}^2_{\alpha} \hat{A}^{2}_{\alpha} + {\varphi}^*_{\alpha}
{\varphi}_{\alpha} (2 \hat{A}^{\dagger}_{\alpha} \hat{A}_{\alpha}
+ 1) +  {\varphi}^{*2}_{\alpha} \hat{A}^{\dagger 2}_{\alpha} ] +
\frac{m_{\rm B}^2}{2} \phi_c^2
 + \frac{\lambda_{\rm B}}{4!} \phi_c^4 \nonumber\\
 && - \frac{\lambda_{\rm B} \hbar^2}{8} \Biggl(\sum_{\alpha}
 \varphi_{\alpha}^* \varphi_{\alpha}\Biggr)^2,
\end{eqnarray}
and
\begin{eqnarray}
\hat{H}_{\rm  P} &=& \frac{\lambda_{\rm B} \hbar^2}{4!}\Biggl\{\sum_{\alpha}
\sum_{k = 0}^{4} {4 \choose k}
{\varphi}^{*(4-k)}_{\alpha} {\varphi}^k_{\alpha} \hat{A}^{\dagger
(4-k) }_{\alpha} \hat{A}^k_{\alpha} \nonumber\\ &&+ \sum_{\alpha \neq \beta}
3 [(\dot{\varphi}^{2}_{\alpha} + \bar{\omega}^2_{\alpha}
{\varphi}^{2}_{\alpha}) \hat{A}^{2}_{\alpha} + 2
(\dot{\varphi}^*_{\alpha} \dot{\varphi}_{\alpha} +
\omega^2_{\alpha} {\varphi}^*_{\alpha} {\varphi}_{\alpha} )
\hat{A}^{\dagger}_{\alpha} \hat{A}_{\alpha} +
(\dot{\varphi}^{*2}_{\alpha} + \bar{\omega}^2_{\alpha}
{\varphi}^{*2}_{\alpha} )\hat{A}^{\dagger 2}_{\alpha}]
\nonumber\\ && \times [\dot{\varphi}^{2}_{\beta} + \omega^2_{\beta}
v^{2}_{\beta}) \hat{A}^{2}_{\beta} + 2
(\dot{\varphi}^*_{\beta} \dot{\varphi}_{\beta} +
\bar{\omega}^2_{\beta} v^*_{\beta} v_{\beta} )
\hat{A}^{\dagger}_{\beta} \hat{A}_{\beta} +
(\dot{\varphi}^{*2}_{\beta} + \bar{\omega}^2_{\beta}
{\varphi}^{*2}_{\beta} ) \hat{A}^{\dagger 2}_{\beta}] \Biggr\}.
\end{eqnarray}
The strategy of the GEP and LvN methods is to exactly solve the
quadratic part $\hat{H}_{\rm G}$ and perturbatively treat the
quartic part $\hat{H}_{\rm P}$. In the GEP method, the GEP is
obtained by minimizing the free energy with respect to a variable
frequency, which consequently determines $\varphi_{\alpha}$. In
the LvN method, the requirement that the creation and annihilation
operators (\ref{ln op}) satisfy the LvN equation (\ref{ln eq})
leads to the mean-field equations for $\varphi_{\alpha}$.

In the GEP method for a static system with time-independent
coupling parameters \cite{chang,stevenson}, one minimizes the
vacuum energy with respect to a trial Gaussian state of the form
\begin{equation}
\Psi_{\Omega_{\alpha}} (\phi_{\alpha}) =
\Biggl(\frac{\Omega_{\alpha}}{\pi} \Biggr)^{1/4} e^{-\frac{i}{2}
\Omega_{\alpha} t} \exp \Biggl[ - \frac{\Omega_{\alpha}}{2 \hbar}
\phi_{\alpha}^2 \Biggr], \label{omega gaus}
\end{equation}
where $\Omega_{\alpha}$ are free parameters. The variational
parameters $\Omega_{\alpha}$ are time-independent for the static
system with time-independent coupling parameters, but depend on
time for a nonequilibrium system with time-dependent coupling
parameters. The time-dependent variational principle by Dirac
\cite{dirac} is required for the time-dependent system, to which
we shall apply the LvN method in this paper. The time-dependent
phase factor in Eq. (\ref{omega gaus}) has been inserted to
satisfy the time-dependent Schr\"{o}dinger equation. Indeed, the
Gaussian state (\ref{omega gaus}) is the vacuum state annihilated
by $\hat{A}$ in Eq. (\ref{ln op}), which can be shown by choosing
the solution
\begin{equation}
\varphi_{\alpha} = \frac{1}{\sqrt{2 \Omega_{\alpha}}} e^{- i
\Omega_{\alpha} t}, \label{sol1}
\end{equation}
and writing Eq. (\ref{ln op}) in the canonical form
\begin{eqnarray}
\hat{A}^{\dagger}_{\alpha} (t) = e^{- i \Omega_{\alpha} t} \Biggl[
\sqrt{\frac{\Omega_{\alpha}}{2 \hbar}} \hat{\phi}_{\alpha} - i
\sqrt{\frac{1}{2 \hbar \Omega_{\alpha}}} \hat{\pi}_{\alpha}
\Biggr], \quad \hat{A}_{\alpha} (t) = e^{i \Omega_{\alpha} t}
\Biggl[ \sqrt{\frac{\Omega_{\alpha}}{2 \hbar}} \hat{\phi}_{\alpha}
+ i \sqrt{\frac{1}{2 \hbar \Omega_{\alpha}}} \hat{\pi}_{\alpha}
\Biggr].
\end{eqnarray}
Then the Hamiltonian leads to the Gaussian effective potential
\begin{equation}
V_{\rm G} = \langle \hat{H} \rangle_{\rm G} = \frac{\hbar}{4}
\sum_{\alpha} \Biggl(\Omega_{\alpha} +
\frac{\bar{\omega}^2_{\alpha}}{\Omega_{\alpha}} \Biggr) +
\frac{\lambda_{\rm B} \hbar^2}{8} \Biggl(\sum_{\alpha} \frac{1}{2
\Omega_{\alpha}} \Biggr)^2 + \frac{m_{\rm B}^2}{2} \phi_c^2 +
\frac{\lambda_{\rm B}}{4!} \phi_c^4. \label{gep}
\end{equation}
The minimization of the Gaussian effective potential with respect
to $\Omega_{\alpha}$ leads to the gap equation
\begin{eqnarray}
\Omega_{\alpha}^2 &=& \bar{\omega}_{\alpha}^2 + \frac{\lambda_{\rm
B} \hbar}{2} \sum_{\beta} \frac{1}{2 \Omega_{\beta}} \nonumber\\
&=& {\bf k}^2 + \mu^2 (\phi_c), \label{gap}
\end{eqnarray}
where
\begin{equation}
\mu^2 (\phi_c) = m_{\rm B}^2 + \frac{\lambda_{\rm B}}{2} \phi_c^2
+ \frac{\lambda_{\rm B} \hbar}{2} \sum_{\beta} \frac{1}{2
\Omega_{\beta}}.
\end{equation}
Now the Gaussian effective potential takes the form
\begin{equation}
V_{\rm G} = \frac{m_{\rm B}^2}{2} \phi_c^2 + \frac{\lambda_{\rm
B}}{4!} \phi_c^4 + \hbar I_1 (\mu) - \frac{\lambda_{\rm B}
\hbar^2}{8} I_0^2 (\mu),
\end{equation}
where
\begin{equation}
I_n (\mu) = \sum_{\alpha} \frac{1}{2 \Omega^{1-2n}_{\alpha}} =
\frac{1}{2}\int \frac{d^3 {\bf k}}{(2 \pi)^{3/2}}
\frac{1}{\Omega_{\bf k}^{1-2n}}. \label{gep int}
\end{equation}

In the GEP method the renormalization of coupling parameters
(constants) are prescribed by matching the coefficients of the
effective potential with renormalized ones. Following Ref.
\cite{stevenson}, the renormalized mass in the Gaussian state, a
symmetric state around $\phi_c = 0$, is given by
\begin{eqnarray}
m_{\rm R}^2 &=& \frac{d^2 V_{\rm G}}{d\phi_c^2} \Biggl|_{\phi_c =
0} \nonumber\\ &=& m_{\rm B}^2 + \frac{\lambda_{\rm B} \hbar}{2}
I_{0}(\mu_0). \label{ren mass}
\end{eqnarray}
Similarly, the renormalized coupling constant is found to be given
by
\begin{eqnarray}
\lambda_{\rm R} &=& \frac{d^4 V_{\rm
G}}{d\phi_{c}^4}\Biggl|_{\phi_c = 0} \nonumber\\ &=& \lambda_{\rm
B} \frac{[1 - \frac{\lambda_{\rm B} \hbar}{2}I_{-1}(\mu)]} {[1 +
\frac{\lambda_{\rm B} \hbar}{4}I_{-1}(\mu)]}. \label{ren coup}
\end{eqnarray}
The case of particularly relevance to our model with a spontaneous
symmetry breaking is when $\lambda_{\rm B}$ is positive and finite
\cite{stevenson}. The other cases corresponding to different
values of $\lambda_{\rm B}$ have also been discussed in detail
\cite{stevenson}. In our model the time-dependent mass term
spontaneously breaks the symmetry during the quenched phase
transition, which may be regarded as a consequence of the
interaction with an environment.

On the other hand, in the LvN method, the operators (\ref{ln op})
are required to satisfy the LvN equation (\ref{ln eq}) for the
truncated Hamiltonian $\hat{H}_{\rm G}$.  This leads to the
mean-field equation for each mode
\begin{equation}
\ddot{\varphi}_{\alpha} + \Biggl[\bar{\omega}^2_{\alpha} +
\frac{\lambda_{\rm B} \hbar}{2} \sum_{\beta}
({\varphi_{\beta}^* {\varphi}_{\beta}}) \Biggr] {\varphi}_{\alpha} = 0.
\label{cl eq}
\end{equation}
The LvN method allows even time-dependent coupling parameters as
long as Eq. (\ref{cl eq}) is satisfied. Further, the operators
(\ref{ln op}) are endowed with the standard commutators
\begin{equation}
[ \hat{A}_{\alpha} (t), \hat{A}_{\beta}^{\dagger} (t) ] =
\delta_{\alpha, \beta},
\end{equation}
which are guaranteed by the Wronskian conditions
\begin{equation}
\dot{\varphi}_{\alpha}^* (t) \varphi_{\alpha} (t) -
\varphi_{\alpha}^* (t) \dot{\varphi}_{\alpha} (t)  = i.
\label{wron}
\end{equation}
The Fock space for each mode is constructed from the
time-dependent creation and annihilation operators,
$\hat{A}^{\dagger}_{\alpha} (t)$ and $\hat{A}_{\alpha} (t)$. The
Gaussian vacuum state of $\alpha$th mode, which is an
approximation to the true vacuum state, is annihilated by
$\hat{A}_{\alpha} (t)$:
\begin{equation}
\hat{A}_{\alpha} (t) \vert 0_{\alpha}, t \rangle_{\rm G} = 0,
\end{equation}
and has the coordinate representation
\begin{equation}
\Psi_{0_{\alpha}} (\phi_{\alpha}, t) = \frac{1}{(2 \pi \hbar
\varphi_{\alpha}^* \varphi_{\alpha})^{1/4}}
\Biggl(\frac{\varphi_{\alpha}}{\varphi_{\alpha}^*} \Biggr)^{1/2}
\exp \Biggl[\frac{i}{2 \hbar}
\frac{\dot{\varphi}_{\alpha}^*}{\varphi_{\alpha}^*}
\phi_{\alpha}^2 \Biggr]. \label{gaus vac}
\end{equation}
Similarly, the $n$th excited state is obtained by applying
$\hat{A}_{\alpha}^{\dagger} (t)$ $n$ times:
\begin{equation}
\vert n_{\alpha}, t \rangle = \frac{1}{\sqrt{n_{\alpha}!}} \Bigl(
\hat{A}_{\alpha}^{\dagger} (t) \Bigr)^{n_{\alpha}} \vert
0_{\alpha}, t \rangle,
\end{equation}
and, in the coordinate representation, is given by
\begin{equation}
\Psi_{n_{\alpha}} (\phi_{\alpha}, t) = \frac{1}{(( 2 \hbar)^n
n!)^{1/2}} \frac{1}{(2 \pi \hbar \varphi_{\alpha}^*
\varphi_{\alpha})^{1/4}}
\Biggl(\frac{\varphi_{\alpha}}{\varphi_{\alpha}^*} \Biggr)^{n +
1/2} H_n \Biggl(\frac{\phi_{\alpha}^2}{\sqrt{2 \hbar
\varphi_{\alpha}^* \varphi_{\alpha}}} \Biggr) \exp
\Biggl[\frac{i}{2 \hbar}
\frac{\dot{\varphi}_{\alpha}^*}{\varphi_{\alpha}^*}
\phi_{\alpha}^2 \Biggr], \label{exc st}
\end{equation}
where $H_n$ is the Hermite polynomial.

A few comments are in order. The quantum states (\ref{gaus vac})
and (\ref{exc st}) in general depend on time, except for trivial
time-dependent phase factors, whenever the coupling parameter
$m_{\rm B}$ or $\lambda_{\rm B}$ depends on time. In fact, these
states constitute the time-dependent Fock space. The Gaussian
vacuum state of the field $\Phi$ is the product of Gaussian state
of each mode:
\begin{equation}
\vert 0, t \rangle_{\rm G} = \prod_{\alpha} \vert 0_{\alpha}, t
\rangle_{\rm G}.
\end{equation}
Similarly, the excited state of the field is the product of each
mode state, at least one state being excited. This Fock
representation of quantum field is unitarily inequivalent because
all excited states of the field are not countable \cite{umezawa}.
Further, there is another feature of the nonequilibrium field that
is absent from the static quantum field. In spite of the fact that
each Fock space at two different times is unitary equivalent
through a Bogoliubov transformation, the Gaussian vacuum states of
the field are orthogonal to each other at two different times due
to infinite number of modes:
\begin{equation}
{}_{\rm G}\langle 0, t' \vert 0, t \rangle_{\rm G} = 0, \quad ( t'
\neq t).
\end{equation}
Now the task in the LvN method is to solve the mean-field equation
(\ref{cl eq}). For the static system the complex function
(\ref{sol1}) is a solution to Eq. (\ref{cl eq}) if
$\Omega_{\alpha}$ satisfies the gap equation (\ref{gap}). This
shows the equivalence between the LvN and GEP methods for at least
static case. Also it follows that the mean-field equations
(\ref{cl eq}) for symmetric quantum states with $\phi_c = 0$, upon
mass renormalization (\ref{ren mass}), become renormalized
\begin{equation}
\ddot{\varphi}_{{\rm R} \alpha} + (m_{\rm R}^2 + {\bf
k}^2){\varphi}_{{\rm R} \alpha} = 0, \label{ren-cl eq}
\end{equation}
where the renormalized frequency is
\begin{eqnarray}
\Omega^2_{{\rm R} \alpha} &=& \Omega^2_{\alpha} \Bigl|_{\phi_c =
0} \nonumber\\ &=& {\bf k}^2 + m_{\rm R}^2. \label{gep ren}
\end{eqnarray}
Therefore, a simple renormalized solution is given by
\begin{equation}
\varphi_{{\rm R} \alpha} = \frac{1}{\sqrt{2 \Omega_{{\rm R}
\alpha}}} e^{- i \Omega_{{\rm R} \alpha} t}.
\end{equation}
This implies that the renormalized Gaussian vacuum state
(\ref{gaus vac}) and its excited states (\ref{exc st}) for the
static Hamiltonian (\ref{free ham}) indeed describe the free
theory with the renormalized mass.

We now compare the Gaussian effective potential (\ref{gep}) with
the conventional perturbation theory. This comparison provides a
direct interpretation of the results of the MSPT method, which
coincide with the results of GEP method to first order of
$\lambda_{\rm B} \hbar$, as will be shown in Secs. IV and V. The
renormalized mass is particularly of interest from the point of
view of instability during the phase transition. The conventional
perturbation theory is based on the solution
\begin{equation}
\varphi^{(0)}_{\alpha} = \frac{1}{\sqrt{2 \bar{\omega}_{\alpha}}}
e^{- i \bar{\omega}_{\alpha} t}. \label{sec sol}
\end{equation}
It is a solution to Eq. (\ref{cl eq}) without the self-interaction
term $(\lambda_{\rm B} = 0)$ and leads to a different Gaussian
vacuum state (\ref{gaus vac}), which is denoted as $\vert 0, t
\rangle_{(0)}$. So one obtains another effective potential
\begin{equation}
V_{(0)} = \langle \hat{H} \rangle_{(0)} = \frac{\hbar}{2}
\sum_{\alpha} \bar{\omega}_{\alpha} + \frac{\lambda_{\rm B}
\hbar^2}{8} \Biggl(\sum_{\alpha} \frac{1}{2 \bar{\omega}_{\alpha}}
\Biggr)^2 + \frac{m_{\rm B}^2}{2} \phi_c^2 + \frac{\lambda_{\rm
B}}{4!} \phi_c^4. \label{sec pot}
\end{equation}
The renormalized mass to order of $(\lambda_{\rm B} \hbar)^2$ is
given by
\begin{eqnarray}
m^2_{{\rm R} (0)} &=& \frac{d^2
V_{(0)}}{d\phi_{c}^2}\Biggl|_{\phi_c = 0} \nonumber\\ &=&  m_{\rm
B}^2 + \frac{\lambda_{\rm B} \hbar}{2} J_{0} - \frac{(\lambda_{\rm
B} \hbar)^2}{8} J_0 J_{-1}, \label{s mass}
\end{eqnarray}
where
\begin{eqnarray}
J_n &=& \sum_{\alpha} \frac{1}{2 {\omega}^{1-2n}_{\alpha}} = \int
\frac{d^3 {\bf k}}{(2 \pi)^{3/2}} \frac{1}{{\omega}^{1-2n}_{\bf
k}}, \nonumber\\ \omega_{\alpha}^2 &=& m^2_{\rm B} + {\bf k}^2.
\end{eqnarray}
To compare Eq. (\ref{s mass}) with the GEP results, (\ref{gep})
and (\ref{gap}), we first expand Eq. (\ref{gap}) as
\begin{equation}
\Omega_{\alpha} (\mu_0) = {\omega}_{\alpha} + \frac{\lambda_{\rm
B} \hbar}{4 {\omega}_{\alpha}} - \frac{(\lambda_{\rm B}
\hbar)^2}{8 {\omega}_{\alpha}} J_{0} J_{-1} - \frac{(\lambda_{\rm
B} \hbar)^2}{32 {\omega}_{\alpha}^3} J_{0}^2 + {\cal O}
(\lambda_{\rm B} \hbar)^3, \label{freq ex}
\end{equation}
and then obtain
\begin{eqnarray}
I_0 (\mu_0) &=& J_0 - \frac{\lambda_{\rm B} \hbar}{4} J_0 J_{-1} +
{\cal O} (\lambda_{\rm B} \hbar)^2, \nonumber\\ I_{-1} (\mu_0) &=&
J_{-1} - \frac{3\lambda_{\rm B} \hbar}{4} J_0 J_{-2} + {\cal O}
(\lambda_{\rm B} \hbar)^2.
\end{eqnarray}
Hence, the renormalized mass (\ref{ren mass})
\begin{equation}
m_{\rm R}^2 = m_{\rm B}^2 + \frac{\lambda_{\rm B} \hbar}{2} J_0 -
\frac{(\lambda_{\rm B} \hbar)^2}{8} J_0 J_{-1} + {\cal O}
(\lambda_{\rm B} \hbar)^3, \label{ren mass ex}
\end{equation}
coincides with Eq. (\ref{s mass}) up to $(\lambda \hbar)^2$. All
higher order terms in the expansion of $m^2_{\rm R}$ and
$\lambda_{\rm R}$ are a consequence of summing over daisy and
superdaisy diagrams in GEP.

Finally, we compare the LvN method with the time-dependent
Hartree-Fock method. We divide the field and the momentum into a
classical background field and a quantum fluctuation:
\begin{equation}
\Phi (t, {\bf X}) = \phi_c + \Phi_q (t, {\bf x}), \quad
\Pi (t, {\bf X}) = \pi_c + \Pi_q (t, {\bf x}).
\end{equation}
The Gaussian state has not only $\langle \hat{\Phi} \rangle =
\phi_c$ but also $\langle \hat{\Pi} \rangle = \pi_c$. The quantum
fluctuations have the zero vacuum expectation value, $\langle
\hat{\Phi}_q \rangle = 0$ and $\langle \hat{\Pi}_q \rangle = 0$.
In the Hartree-Fock method, $\hat{\Phi}_q^3$ is replaced by $3
\langle \hat{\Phi}^2_q \rangle \hat{\Phi}_q$. Then the field
equation for quantum fluctuation is given by
\begin{equation}
\partial_{\mu} \partial^{\mu} \hat{\Phi}_q + \Biggl( m_{\rm B}^2
+  \frac{\lambda_{\rm B} }{2} \phi_c^2 + \frac{\lambda_{\rm B}}{2}
\langle \hat{\Phi}^2_q \rangle \Biggr) \hat{\Phi}_q = 0,
\label{ren-fl eq}
\end{equation}
and for classical field by
\begin{equation}
\partial_{\mu} \partial^{\mu} \phi_c + \Biggl(m_{\rm B}^2 +
\frac{\lambda_{\rm B}}{2} \langle \hat{\Phi}^2_q \rangle \Biggr)
\phi_c + \frac{\lambda_{\rm B}}{3!} \phi_c^3 = 0. \label{ren-cl
eq2}
\end{equation}
To remove the infinite term
\begin{equation}
\langle \hat{\Phi}_q^2 \rangle = \hbar \sum_{\alpha} \varphi_{\alpha}^*
\varphi_{\alpha} = \hbar I_0 (\mu_0),
\end{equation}
we add counter terms $(\delta m^2/2) \Phi^2$ and $(\delta
\lambda/4!) \Phi^4$, which is equivalent to writing the bare
coupling parameters as
\begin{equation}
m_{\rm B}^2 = m_{\rm R}^2 + \delta m^2, \quad \lambda_{\rm B} =
\lambda_{\rm R} + \delta \lambda.
\end{equation}
By choosing the counter mass term as $\delta m^2 = - (\lambda_{\rm
B} \hbar/2) I_0 (\mu_0)$, we obtain from Eq. (\ref{ren-fl eq}) the
renormalized field equation for $\phi_c =0$
\begin{equation}
(\partial_{\mu} \partial^{\mu}  + m^2_{\rm R}) \hat{\Phi}_{{\rm R}
q} = 0. \label{ren-fl eq2}
\end{equation}
Note that the renormalized mass obtained by adding the mass
counter term is the same as Eq. (\ref{ren mass}) in the GEP or the
LvN method and the Fourier transform of Eq. (\ref{ren-fl eq2})
would be the operator counterpart of the renormalized mean-field
equation (\ref{ren-cl eq}).

\section{Analysis of the Coupled Quartic Oscillator Model using MSPT Method}

The multiple scale perturbation theory (MSPT) was introduced to
find solutions to nonlinear systems with a perturbatively small
coupling constant, say $\lambda \hbar$ \cite{bender,bender2}. The
nonlinear systems exhibit distinct characteristic behaviour on
different time scales determined by $\lambda \hbar$. A
conventional perturbation theory leads to secular (boundlessly
growing) terms due to the resonant coupling between a lower order
and the next leading order. However, the MSPT method provides a
systematic technique of eliminating the secular terms and, as a
consequence, gives rise to the solution with an effective
frequency, which can be obtained by summing the most secular terms
to all orders of $\lambda \hbar$ in the conventional perturbation
theory \cite{bender2}. Even the lowest order MSPT solution is
valid over a time scale $1/(\lambda\hbar)$. In this section we
apply the MSPT method to analyze the system of two coupled quartic
oscillators. It is a useful toy model for getting some insight
into the phenomenon of mode mixing that one encounters in the
study of nonequilibrium quantum fields during a spontaneous
symmetry breaking phase transition. The MSPT was applied to a
single classical as well as a quantum anharmonic oscillator in
Ref. \cite{bender2}. In the latter case, the technique involves
solving a complicated set of coupled operator equations. However,
in the LvN method, the auxiliary equations for the Gaussian state
or its excited states are the classical mean-field type equations
or the linearized Hartree-Fock equations, which can be more easily
tackled. As mentioned earlier (see also \cite{kim-lee2}), these
coupled equations are a direct consequence of the fact that the
creation and annihilation operators given in Eq. (11) satisfy the
LvN equation.

The model to be studied in this section is described by the Hamiltonian
\begin{equation}
H = \frac{p_1^2}{2} + \frac{\omega_1^2}{2} q_1^2 +  \frac{p_2^2}{2}
+ \frac{\omega_2^2}{2} q_2^2 + \frac{\lambda}{4!} \Biggl(q_1^4
+ 6 q_1^2 q_2^2 + q_2^4 \Biggr). \label{mod ham}
\end{equation}
The Hamiltonian (\ref{mod ham}) is obtained by taking two typical
modes of scalar field from the field Hamiltonian (\ref{ham-eff}).
The most physically interesting case is the coupling between the
hard and soft modes. The approximate Gaussian state for the
Hamiltonian (\ref{mod ham}) with $\langle \hat{q}_1 \rangle = 0 =
\langle \hat{q}_2 \rangle$ is given by the product
\begin{equation}
\noindent \Psi_0 (q_1, q_2) = \Psi_1 (q_1) \Psi_2 (q_2)
\end{equation}
of the Gaussian state for each mode:
\begin{equation}
\Psi_i (q_i) = \frac{1}{(2 \pi \hbar v_i^* v_i)^{1/4}}
\Biggl(\frac{v_i}{v_i^*} \Biggr)^{1/2} \exp\Biggl[\frac{i}{2
\hbar} \frac{\dot{v}_i^*}{v_i^*} q_i^2 \Biggr]. \label{mod gaus}
\end{equation}
The $v_i, (i = 1, 2),$ satisfy the coupled equations
\begin{eqnarray}
\ddot{v}_1 + \omega_1^2 v_1 + \frac{\lambda \hbar}{2} (v_1^* v_1 + v_2^* v_2
 ) v_1 = 0, \nonumber\\
\ddot{v}_2 + \omega_2^2 v_2 + \frac{\lambda \hbar}{2} (v_1^* v_1 +
v_2^* v_2
 ) v_2 =  0, \label{mod eq1}
\end{eqnarray}
together with the Wronskian conditions
\begin{equation}
\dot{v}_i^* v_i - \dot{v}_i v_i^* = i. \label{wron2}
\end{equation}
Note that $(\lambda \hbar)$ in Eq. (\ref{mod eq1}) is the
nonlinear coupling constant. As each time scale of order $(\lambda
\hbar)^n t \sim {\cal O}(1)$ with $n = 1, 2, \cdots,$ shows a
different characteristic behavior of the solution, we introduce a
multiple of different time scales
\begin{equation}
\tau_{(1)} = (\lambda \hbar) t, \cdots, \tau_{(n)} = (\lambda
\hbar)^n t, \cdots.
\end{equation}
The solution to Eq. (\ref{mod eq1}) is expanded in a series of
$\lambda \hbar$ as
\begin{equation}
v_i = \sum_{n = 0} (\lambda \hbar)^n
v_i^{(n)} (t, \tau_{(1)}, \cdots, \tau_{(n)},
\cdots), \label{mod ser}
\end{equation}
where $v_i^{(n)}$ is of the order of unity. The time derivative is now
given by
\begin{eqnarray}
\frac{\partial^2 v_i}{\partial t^2} &=&
\frac{\partial^2 v_i^{(0)}}{\partial t^2}  + 2
\frac{\partial^2 v_i^{(0)}}{\partial t \partial \tau_{(1)}}
\Biggl(\frac{\partial \tau_{(1)}}{\partial t} \Biggr)
+ \frac{\partial^2 v_i^{(0)}}{\partial \tau_{(1)}^2}
\Biggl(\frac{\partial \tau_{(1)}}{\partial t} \Biggr)^2 +
 2 \frac{\partial^2 v_i^{(0)}}{\partial t \partial \tau_{(2)}}
\Biggl(\frac{\partial \tau_{(2)}}{\partial t} \Biggr) + \cdots
\nonumber\\ && + \lambda \hbar \Biggl\{\frac{\partial^2
v_i^{(1)}}{\partial t^2} + 2 \frac{\partial^2 v_i^{(1)}}{\partial
t \partial \tau_{(1)}} \Biggl(\frac{\partial \tau_{(1)}}{\partial
t} \Biggr) + \frac{\partial^2 v_i^{(1)}}{\partial \tau^2_{(1)}}
\Biggl(\frac{\partial \tau_{(1)}}{\partial t} \Biggr)^2 + \cdots
\Biggr\} \nonumber\\ && +  (\lambda \hbar)^2 \Biggl\{
\frac{\partial^2 v_i^{(2)}}{\partial t^2} + 2 \frac{\partial^2
v_i^{(2)}}{\partial t \partial \tau_{(1)}} \Biggl(\frac{\partial
\tau_{(1)}}{\partial t} \Biggr) + \frac{\partial^2
v_i^{(2)}}{\partial \tau^2_{(1)}} \Biggl(\frac{\partial
\tau_{(1)}}{\partial t} \Biggr)^2 + \cdots \Biggr\} + \cdots
\label{mod eq2}
\end{eqnarray}
To zeroth order $(\lambda \hbar)^0$, $v_i^{(0)}$ satisfies the
simple oscillator equation
\begin{equation}
\ddot{v}_i^{(0)} + \omega^2_i v_i^{(0)} = 0. \label{zero eq}
\end{equation}

First, we consider the case $\omega_i^2 \geq 0$, which is the
quantum mechanical analog of the coupling between two hard modes.
As $\tau \equiv \tau_{(1)}$ is a time scale independent of $t$,
one looks for the zeroth order solution to Eq. (\ref{zero eq}) of
the form
\begin{equation}
v_i^{(0)} = A_i (\tau) e^{i \omega_i t}
+ B_i (\tau) e^{- i \omega_i t}. \label{mod sol1}
\end{equation}
Collecting terms of order $(\lambda \hbar)$ from Eq. (\ref{mod
eq2}), one obtains the nonlinear equation
\begin{equation}
\Biggl\{\frac{\partial^2 v_i^{(1)}}{\partial t^2}  + \omega_i^2
v_i^{(1)} \Biggr \} + \Biggl\{2 \frac{\partial^2
v_i^{(0)}}{\partial  t \partial \tau} + \frac{1}{2} \Biggl(
v_1^{(0)*} v_1^{(0)} + v_2^{(0)*}v_2^{(0)} \Biggr) v_i^{(0)}
\Biggr\} = 0. \label{mod eq3}
\end{equation}
The secular terms, which are proportional to $e^{\pm i \omega_i
t}$ in the second curly bracket of Eq. (\ref{mod eq3}), should be
eliminated to prevent $v_i^{(1)}$ from growing boundlessly in
time. This requires the coefficients to satisfy
\begin{eqnarray}
\frac{\partial A_i}{\partial \tau} &=& \frac{i}{4 \omega_i}
\Biggl\{B_i^* B_i + \sum_{j = 1,2} A_j^* A_j + B_j^* B_j \Biggr\}
A_i, \nonumber\\ \frac{\partial B_i}{\partial \tau} &=& \frac{-
i}{4 \omega_i} \Biggl\{A_i^* A_i + \sum_{j = 1, 2} A_j^* A_j +
B_j^* B_j \Biggr\} B_i. \label{mod eq4}
\end{eqnarray}
Equation (\ref{mod eq4}) together with its complex conjugate for
the coefficients $A_i^* (\tau_{(1)})$ and $B_i^* (\tau_{(1)})$,
determines the initial values
\begin{equation}
A_i^* (\tau) A_i (\tau) = A_i^{*}(0) A_i (0), \quad B_i^* (\tau)
B_i (\tau) =  B_i^*(0) B_i (0).
\end{equation}
Now the solutions to Eq. (\ref{mod eq4}) are given by
\begin{eqnarray}
A_i (\tau) &=& A_i (0) \exp \Biggl[ \frac{i \tau}{4 \omega_i }
\Biggl\{B_i^* (0) B_i (0) +  \sum_{j = 1, 2} (A_j^* (0) A_j (0) +
B_j^* (0) B_j (0)) \Biggr\} \Biggr], \nonumber\\ B_i (\tau) &=&
B_i (0) \exp \Biggl[ \frac{- i \tau}{4 \omega_i }\Biggl\{A_i^* (0)
A_i (0) + \sum_{j = 1, 2} (A_j^* (0) A_j (0) + B_j^* (0) B_j (0))
\Biggr\} \Biggr]. \label{mod sol2}
\end{eqnarray}
At order $(\lambda \hbar)^0$, the initial Gaussian state is given
by
\begin{equation}
A_i (0) = 0, \quad B_i (0) = \frac{1}{\sqrt{2
\omega_i}}. \label{mod data}
\end{equation}
Therefore, the solution for $v^{(0)}_i$ becomes
\begin{equation}
v^{(0)}_i = \frac{1}{ \sqrt{2 \omega_i}} e^{- i \Omega_i^{(-)} t},
\label{mod sol7}
\end{equation}
where each mode has a new shifted frequency
\begin{equation}
\Omega^{(-)}_i = \omega_i + \frac{\lambda \hbar}{4 \omega_i}
\sum_{j = 1, 2} \frac{1}{2 \omega_j}. \label{mod freq}
\end{equation}
The second term can be obtained by summing the most secular terms
to all orders in the conventional perturbation theory. As each
$v_i^{(0)}$ satisfies the linear equation (\ref{zero eq}), we
choose the coefficient to satisfy the Wronskian condition
(\ref{wron2})
\begin{equation}
v_i = \frac{1}{ \sqrt{2 \Omega_i^{(-)}}} e^{- i \Omega_i^{(-)} t}.
\label{first sol}
\end{equation}
Note that the solution (\ref{first sol}), when expanded in power
of $\lambda \hbar$, yields the first order solution $v_i^{(1)}$,
as it has the correct frequency and coefficient up to that order.
By comparing Eq. (\ref{mod freq}) with the gap equation
\begin{equation}
\Omega_i = \Biggl[\omega_i^2 + \frac{\lambda \hbar}{2} \sum_{j =
1, 2} \frac{1}{2 \Omega_j} \Biggr]^{1/2}, \label{gaus freq}
\end{equation}
obtained from the exact solution to Eq. (\ref{mod eq1})
\begin{equation}
v_i = \frac{1}{\sqrt{2 \Omega_i}} e^{- i \Omega_i t},
\end{equation}
it can be shown that $\Omega_i^{(-)}$ is the binomial expansion of
$\Omega_i$ to order of $\lambda \hbar$. Hence the lowest order
MSPT solution gives the correct frequency or the energy to order
$(\lambda \hbar)$.

We now consider the more interesting case of one hard and one soft
mode. To mimic the interaction between the hard and soft modes
after a sudden quench, one frequency squared, say $\omega_2^2$,
will instantaneously change the sign from positive to negative at
$t=0$, i.e., $\omega_{1}^2 \rightarrow \tilde{\omega}_1^2$ and
$\omega_{2}^2 \rightarrow - \tilde{\omega}_2^2$. Then Eq.
(\ref{mod ham}) for $t
> 0$ takes the form
\begin{equation}
\tilde{H} = \frac{p_1^2}{2} + \frac{\tilde{\omega}_1^2}{2} q_1^2 +
\frac{p_2^2}{2} - \frac{\tilde{\omega}_2^2}{2} q_2^2 +
\frac{\lambda}{4!} \bigl(q_1^4 + 6 q_1^2 q_2^2 + q_2^4 \bigr),
\label{mod ham2}
\end{equation}
exactly the Hamiltonian for the interaction between a hard mode
and a soft mode in $\Phi^4$-theory which will be discussed in the
next section. The equations for the auxiliary variables
$\tilde{v}_{1}$ and $\tilde{v}_{2}$ to ${\cal{O}}((\lambda
\hbar)^{0})$ become
\begin{eqnarray}
\ddot{\tilde{v}}_1^{(0)} + \tilde{\omega}^2_1 \tilde{v}_1^{(0)}
&=& 0, \nonumber\\ \noindent \ddot{\tilde{v}}_2^{(0)} -
\tilde{\omega}^2_2 \tilde{v}_2^{(0)} &=& 0. \label{hs eq2}
\end{eqnarray}
As for the former case of two hard modes, $\tau = (\lambda \hbar)
t$ introduces another time scale in addition to $t$, and the
solutions to Eq. (\ref{hs eq2}) take the form
\begin{eqnarray}
\tilde{v}_1^{(0)} &=& \tilde{A} (\tau) e^{i \tilde{\omega}_1 t} +
\tilde{B} (\tau) e^{- i \tilde{\omega}_1 t}, \nonumber\\ \noindent
\tilde{v}_2^{(0)} &=& \tilde{C} (\tau) e^{ \tilde{\omega}_2 t} +
\tilde{D} (\tau) e^{- \tilde{\omega}_2 t}. \label{mod sol3}
\end{eqnarray}
Secular terms appear in equations for $\tilde{v}^{(1)}_i$ due to
the resonant coupling with $\tilde{v}^{(0)}_i$. These are the
terms proportional to $e^{\pm i \tilde{\omega}_{1} t}$ for
$\tilde{v}^{(1)}_1$ and $e^{\pm \tilde{\omega}_{2} t}$ for
$\tilde{v}^{(2)}_2$ and lead to boundlessly growing solutions. By
eliminating these terms, one obtains the equations
\begin{eqnarray}
\frac{\partial \tilde{A}}{\partial \tau} &=& \frac{i}{4
\tilde{\omega}_1} \Biggl\{ \tilde{A}^* \tilde{A} + 2 \tilde{B}^*
\tilde{B} + \tilde{C}^* \tilde{D} + \tilde{C} \tilde{D}^* \Biggr\}
\tilde{A}, \nonumber\\ \frac{\partial \tilde{B}}{\partial \tau}
&=& \frac{- i}{4 \tilde{\omega}_1} \Biggl\{ 2 \tilde{A}^*
\tilde{A} + \tilde{B}^* \tilde{B} + \tilde{C}^* \tilde{D} +
\tilde{C} \tilde{D}^* \Biggr\} \tilde{B}, \nonumber\\
\frac{\partial \tilde{C}}{\partial \tau} &=& \frac{- 1}{4
\tilde{\omega}_2} \Biggl\{ \tilde{A}^* \tilde{A} + \tilde{B}^*
\tilde{B} + 2 \tilde{C}^* \tilde{D} + \tilde{C} \tilde{D}^*
\Biggr\} \tilde{C}, \nonumber\\ \frac{\partial \tilde{D}}{\partial
\tau} &=&  \frac{1}{4 \tilde{\omega}_2} \Biggl\{ \tilde{A}^*
\tilde{A} + \tilde{B}^* \tilde{B} +  \tilde{C}^* \tilde{D} + 2
\tilde{C} \tilde{D}^* \Biggr\} \tilde{D}. \label{mod eq5}
\end{eqnarray}
A similar set of equations involving coefficients $\tilde{A}^*$,
$\tilde{B}^*$, $\tilde{C}^*$, $\tilde{D}^*$, is obtained by
requiring that the secular terms for $\tilde{v}_{1}^{(1)*}$ and
$\tilde{v}_{2}^{(1)*}$ vanish, which is the complex conjugate of
Eq. (\ref{mod eq5}). Then it follows that
\begin{eqnarray}
\tilde{A}^* (\tau) \tilde{A} (\tau) &=& \tilde{A}^* (0) \tilde{A}
(0), \quad \tilde{B}^* (\tau) \tilde{B} (\tau) = \tilde{B}^* (0)
\tilde{B} (0), \nonumber\\ \tilde{C}^* (\tau) \tilde{D} (\tau) &=&
\tilde{C}^* (0) \tilde{D} (0), \quad \tilde{C} (\tau) \tilde{D}^*
(\tau) = \tilde{C} (0) \tilde{D}^* (0). \label{coef eq}
\end{eqnarray}

As the system evolves from an initial Gaussian state for $t < 0$
to another Gaussian state for $t > 0$ after the phase transition,
the Gaussian state (\ref{mod gaus}) should be continuous across
$t= 0$. The continuity of the Gaussian state is equivalent to the
continuity of $v_i$ and $\tilde{v}_i$ at $t=0$, which determines
the coefficients of the zeroth order solutions after the phase
transition in terms of the initial data:
\begin{eqnarray}
\tilde{A}(0) &=& \frac{1}{2 \sqrt{2\omega_1}}\Biggl(1 -
\frac{\omega_1}{\tilde{\omega}_1}\Biggr), \quad \noindent
\tilde{B}(0) = \frac{1}{2 \sqrt{2\omega_1}} \Biggl(1 +
\frac{\omega_1}{\tilde{\omega}_1}\Biggr), \nonumber\\ \tilde{C}_2
(0) &=& \frac{1}{2 \sqrt{2 \omega_2}}\Biggl(1 - i
\frac{{\omega}_2}{\tilde{\omega}_2}\Biggr), \quad \noindent
\tilde{D}_2 (0) = \frac{1}{2 \sqrt{2 \omega_2}}\Biggl(1 + i
\frac{{\omega}_2}{\tilde{\omega}_2}\Biggr). \label{coef eq2}
\end{eqnarray}
By substituting Eq. (\ref{coef eq2}) into Eq. (\ref{coef eq}) and
solving Eq. (\ref{mod eq5}), one finally obtains
\begin{eqnarray}
\tilde{v}_1^{(0)} (t) &=& \tilde{A} (0) e^{i
\tilde{\Omega}^{(+)}_1 t} + \tilde{B} (0) e^{- i
\tilde{\Omega}^{(-)}_1 t}, \nonumber\\ \tilde{v}_2^{(0)} (t) &=&
\tilde{C} (0) e^{\tilde{\Omega}^{(+)}_2 t} + \tilde{D} (0) e^{-
\tilde{\Omega}^{(-)}_2 t}, \label{mod sol9}
\end{eqnarray}
where
\begin{eqnarray}
\tilde{\Omega}_{1}^{(\pm)} &=& \tilde{\omega}_1 + \frac{\lambda
\hbar}{4 \tilde{\omega}_1} \Biggl\{\frac{3}{8 \omega_1}\Biggl(1 +
\frac{\omega_1^2}{\tilde{\omega}_1^2}\Biggr) + \frac{1}{4
\omega_2}\Biggl(1 - \frac{\omega_2^2}{\tilde{\omega}_2^2}\Biggr)
\pm \frac{1}{4\tilde{\omega}_1}\Biggr\}, \nonumber\\
\tilde{\Omega}_{2}^{(\pm)} &=& \tilde{\omega}_2 - \frac{\lambda
\hbar}{4 \tilde{\omega}_2} \Biggl\{\frac{1}{4 \omega_1}\Biggl(1 +
\frac{\omega_1^2}{\tilde{\omega}_1^2}\Biggr) + \frac{3}{8
\omega_2}\Biggl(1 - \frac{\omega_2^2}{\tilde{\omega}_2^2}\Biggr)
\pm \frac{i}{4\tilde{\omega}_2}\Biggr\}.
\end{eqnarray}
By requiring the Wronskian conditions (\ref{wron2}) and matching
the first order solutions (\ref{first sol}), we find the correct
first order coefficients as
\begin{eqnarray}
\tilde{v}_1 (t) &=&
\frac{1}{\sqrt{2\Omega^{(-)}_1}}\Biggl(\frac{\tilde{\Omega}^{(-)}_1
- \Omega^{(-)}_1}{\tilde{\Omega}^{(+)}_1 + \tilde{\Omega}^{(-)}_1
}\Biggr) e^{i \tilde{\Omega}^{(+)}_1 t} + \frac{1}{
\sqrt{2\Omega^{(-)}_1}}\Biggl(\frac{\tilde{\Omega}^{(+)}_1 +
\Omega^{(-)}_1}{\tilde{\Omega}^{(+)}_1 + \tilde{\Omega}^{(-)}_1
}\Biggr) e^{- i \tilde{\Omega}^{(-)}_1 t}, \nonumber\\ \tilde{v}_2
(t) &=&
\frac{1}{\sqrt{2\Omega^{(-)}_2}}\Biggl(\frac{\tilde{\Omega}^{(-)}_1
- i \Omega^{(-)}_1}{\tilde{\Omega}^{(+)}_1 +
\tilde{\Omega}^{(-)}_1 }\Biggr)  e^{\tilde{\Omega}^{(+)}_2 t} +
\frac{1}{\sqrt{2\Omega^{(-)}_2}}\Biggl(\frac{\tilde{\Omega}^{(+)}_1
+i \Omega^{(-)}_1}{\tilde{\Omega}^{(+)}_1 + \tilde{\Omega}^{(-)}_1
}\Biggr)  e^{- \tilde{\Omega}^{(-)}_2 t}. \label{mod sol20}
\end{eqnarray}

A few comments on the interpretation of solutions (\ref{mod sol7})
and (\ref{mod sol20}) are in order.  These solutions are valid
till $t \sim 1/(\lambda \hbar)$. The coupling between the hard
modes just modifies the frequencies (\ref{mod freq}) even after
the phase transition. However, the coupling between the hard and
soft modes after the phase transition exhibits some interesting
features for the both modes. One interesting feature is the
disparity between the positive and negative frequency given by
$\delta \tilde{\Omega}_1 \equiv \tilde{\Omega}_1^{(+)} -
\tilde{\Omega}_1^{(-)} = \lambda/8 \tilde{\omega}_1^2$ for the
hard mode and $\delta \tilde{\Omega}_2 \equiv
\tilde{\Omega}_2^{(+)} - \tilde{\Omega}_2^{(-)} = -
i\lambda/8\tilde{\omega}_2^2$ for the soft mode. This difference
is a consequence of the resonant coupling for the positive and the
negative frequency solution and for the exponentially growing and
decaying solutions. Another interesting feature is the appearance
of an imaginary part of $\tilde{\Omega}_2^{(\pm)}$ for the soft
mode. This implies that $\tilde{v}_2$ begins to oscillate due to
the coupling between the hard and hard modes. The time scale of
oscillation is estimated to be $(16 \omega_2^2)/(\lambda \hbar)$,
which is larger by the factor $16 \omega_2^2$ than the time scale
$1/\lambda \hbar$ for the lowest order solution to be valid. The
period of the oscillation is too long to be observed.

Though the time scale for oscillation is outside the validity
region of the zeroth order solution, the oscillatory behavior
gives more information about the later stage of evolution towards
and oscillation around the true vacuum. The numerical analysis of
the first order solution shows that the early stage of exponential
growth due to instability is eventually terminated by the
back-reaction from nonlinear terms \cite{ksk}. It is the
oscillatory part that determines the absolute width of the
Gaussian state after the field samples the true vacuum. To see how
the oscillatory part to restrict the Gaussian state, we use the
lowest order solution $\tilde{v}_2$, which is the auxiliary
equation for the Gaussian state for the soft mode. The correct
physical interpretation may follow from the Gaussian state
(\ref{mod gaus}):
\begin{eqnarray}
\tilde{\Psi}_2 (q_2, t) =  \frac{1}{(2 \pi \hbar \tilde{v}_2^*
\tilde{v}_2)^{1/4}} \Biggl(\frac{\tilde{v}_2}{\tilde{v}_2^*}
\Biggr)^{1/2} \exp \Biggl[ - \Biggl\{\frac{\lambda}{32
\tilde{\omega}_2^2} + \frac{2\omega_2}{\hbar \tilde{\omega}_2}
{\rm Re}(\tilde{\omega}^{(+)}_2) e^{- 2 {\rm
Re}(\tilde{\omega}^{(+)}_2) t} + \cdots \Biggr\}q^2_2 \nonumber\\
+ i \{\frac{{\rm Re}(\tilde{\omega}^{(+)}_2)}{2 \hbar} (1 - e^{- 4
{\rm Re}(\tilde{\omega}^{(+)}_2) t} + \cdots) \} q_2^2 \Biggr].
\label{gaus dis}
\end{eqnarray}
Immediately after the phase transition, the width (dispersion) of
the Gaussian state (\ref{gaus dis}) is largely determined by the
second term of the real part which originates from the
exponentially growing part of $\tilde{v}_2$. But, as the phase
transition continues, the first term from the imaginary part of
$\tilde{\Omega}_2^{(\pm)}$ is comparable to the second term when
\begin{equation}
e^{- 2 {\rm Re}(\tilde{\omega}^{(+)}_2) t} \approx \frac{\lambda
\hbar}{64 \tilde{\omega}_2^3}. \label{osc time}
\end{equation}
The time scale of Eq. (\ref{osc time}) is outside the validity of
the lowest order solution. Before reaching the time (\ref{osc
time}), the back-reaction $(\lambda \hbar/2) (\tilde{v}_2^{*}
\tilde{v}_2) \tilde{v}_2$ of the self-interaction begins to
dominate over $- \tilde{\omega}_2^2 \tilde{v}_2$. So the more
accurate first order solution oscillates around the true vacuum
after an exponentially growing period and the Gaussian state has
approximately a constant width, like the first term of the real
part. That is, the Gaussian state stops spreading.

\section{MSPT and Renormalized Frequency}

The main advantages of the MSPT method, especially within the
framework of the LvN formalism, was discussed in the previous
section. Nevertheless, it is necessary to make some comments on
this method in the field theoretic context. The process of
eliminating the secular terms at various orders of the MSPT method
is equivalent to summing certain diagrams to all orders in the
conventional perturbation theory. It is precisely because of this
fact that useful information pertaining to the divergence
structure of the theory can be extracted even from the lowest
order solution in the MSPT method. As will be shown below, the
lowest order MSPT solution of a quantum field yields a shifted
(effective) frequency which contains divergent contributions and
the removal of the most divergent term essentially amounts to mass
renormalization.

The requirement that the creation and annihilation operators (11)
satisfy the LvN equation, leads to the mean-field equation of the
auxiliary field variable of each mode
\begin{equation}
\ddot{\varphi}_{\alpha} + \omega^2_{\alpha} \varphi_{\alpha} +
\frac{\lambda_{\rm B} \hbar}{2} \Biggl(\sum_{\beta}
\varphi_{\beta}^* \varphi_{\beta} \Biggr) \varphi_{\alpha} = 0,
\label{mean eq}
\end{equation}
where
\begin{equation}
\omega_{\alpha}^2 = m_{\rm B}^2 + {\bf k}^2.
\end{equation}
The exact solution to Eq. (\ref{mean eq}) for the static system
without phase transition is given by Eq. (\ref{sol1}). Now we
apply the MSPT to solve perturbatively Eq. (\ref{mean eq}) in a
power series of $\lambda_{\rm B} \hbar$ and to find the correct
renormalized frequency. As in Sec. III, introduce a multiple of
time scales
\begin{equation}
\tau_{(1)} = (\lambda_{\rm B} \hbar) t, \cdots, \tau_{(n)} =
(\lambda_{\rm B} \hbar)^n t, \cdots,
\end{equation}
and expand the solution in the power series of $\lambda \hbar$
\begin{equation}
\varphi_{\alpha} = \sum_{n = 0} (\lambda_{\rm B} \hbar)^n
\varphi_{\alpha}^{(n)} (t, \tau_{(1)}, \cdots, \tau_{(n)},
\cdots). \label{mspt ser}
\end{equation}
As we shall find the solution (\ref{mspt ser}) only to the first
order, we denote $\tau_{(1)} = \tau$ and truncate it to that order
\begin{equation}
\varphi_{\alpha} (t)  = \varphi_{\alpha}^{(0)} (t)  +
(\lambda_{\rm B} \hbar) \varphi_{\alpha}^{(1)} (t, \tau).
\end{equation}
The time derivative is now given by
\begin{equation}
\frac{\partial^2  \varphi_{\alpha} (t)}{\partial t^2} =
\frac{\partial^2  \varphi_{\alpha}^{(0)}}{\partial t^2} + 2
\frac{\partial^2  \varphi_{\alpha}^{(0)} (t)}{\partial t \partial \tau}
\Biggl(\frac{\partial \tau}{\partial t} \Biggr) + (\lambda_{\rm B}
\hbar) \frac{\partial^2  \varphi_{\alpha}^{(1)}}{\partial t^2} +
\cdots. \label{mspt eq}
\end{equation}
The terms to order $(\lambda_{\rm B} \hbar)^0$ satisfy the
equation for a simple oscillator
\begin{equation}
\ddot{\varphi}_{\alpha}^{(0)} + \omega^2_{\alpha}
\varphi_{\alpha}^{(0)} = 0.
\end{equation}
The general solution is a superposition of the positive and
negative frequency solutions
\begin{equation}
\varphi_{\alpha}^{(0)} = A_{\alpha} (\tau) e^{i \omega_{\alpha} t}
+ B_{\alpha} (\tau) e^{- i \omega_{\alpha} t}. \label{mspt sol1}
\end{equation}

At the next order the terms of order $(\lambda_{\rm B} \hbar)$  in
Eq. (\ref{mean eq}) obey the equation
\begin{equation}
\Biggl\{\frac{\partial^2 \varphi_{\alpha}^{(1)}}{\partial t^2}  +
\omega_{\alpha}^2 \varphi_{\alpha}^{(1)} \Biggr \} + \Biggl\{2
\frac{\partial^2 \varphi_{\alpha}^{(0)}}{\partial  t \partial \tau}
 + \frac{1}{2} \Biggl(\sum_{\beta}
\varphi_{\beta}^{(0)*} \varphi_{\beta}^{(0)} \Biggr)
\varphi_{\alpha}^{(0)} \Biggr\} = 0. \label{mspt 1}
\end{equation}
As $\varphi^{(1)}_{\alpha}$ has the same natural frequency
$\omega_{\alpha}$ as $\varphi^{(0)}_{\alpha}$, any term
proportional to $e^{\pm i \omega_{\alpha} t}$ in the second curly
bracket is a secular term to $\varphi_{\alpha}^{(1)}$. To avoid
unnecessary growing terms, these secular terms should be
eliminated from the second curly bracket. This requires
\begin{eqnarray}
\frac{\partial  A_{\alpha}}{\partial \tau} &=& \frac{i}{4
\omega_{\alpha}} \Biggl\{ B_{\alpha}^* B_{\alpha}  + \sum_{\beta} A_{\beta}^* A_{\beta} +
B_{\beta}^* B_{\beta} \Biggr\} A_{\alpha}, \nonumber\\
\frac{\partial  B_{\alpha}}{\partial \tau} &=& \frac{- i}{4
\omega_{\alpha}} \Biggl\{A_{\alpha}^* A_{\alpha}  + \sum_{\beta}  A_{\beta}^* A_{\beta} +
B_{\beta}^* B_{\beta} \Biggr\} B_{\alpha}. \label{mspt eq2}
\end{eqnarray}
As the coefficients of $A_{\alpha}$ and $B_{\alpha}$ are real, Eq.
(\ref{mspt eq2}) is unitary, so has the solution
\begin{equation}
A_{\alpha}^* (\tau) A_{\alpha} (\tau) = A_{\alpha}^* (0)
A_{\alpha} (0), \quad B_{\alpha}^* (\tau) B_{\alpha} (\tau) =
B_{\alpha}^* (0) B_{\alpha} (0).
\end{equation}
Hence the solutions are given by
\begin{eqnarray}
A_{\alpha} (\tau) &=& A_{\alpha} (0) \exp \Biggl[ \frac{i \tau}{4
\omega_{\alpha} }\Biggl\{B_{\alpha}^* (0) B_{\alpha} (0)  +
\sum_{\beta} A_{\beta}^* (0) A_{\beta} (0) + B_{\beta}^* (0)
B_{\beta} (0) \Biggr\} \Biggr], \nonumber\\ B_{\alpha} (\tau) &=&
B_{\alpha} (0) \exp \Biggl[\frac{- i \tau}{4 \omega_{\alpha} }
\Biggl\{A_{\alpha}^* (0)A_{\alpha}(0)  +  \sum_{\beta} A_{\beta}^*
(0) A_{\beta} (0) + B_{\beta}^* (0) B_{\beta} (0) \Biggr\}
\Biggr]. \label{mspt sol2}
\end{eqnarray}
As the Gaussian state before the phase transition has the initial
data
\begin{equation}
A_{\alpha} (0) = 0, \quad B_{\alpha} (0) = \frac{1}{\sqrt{2
\omega_{\alpha}}}, \label{init data}
\end{equation}
the solution becomes
\begin{equation}
\varphi_{\alpha}^{(0)} = \frac{1}{ \sqrt{2 \omega_{\alpha}}} \exp
[- i \Omega^{(-)}_{\alpha} t], \label{zero sol}
\end{equation}
where
\begin{equation}
\Omega^{(-)}_{\alpha} = \omega_{\alpha} + \frac{\lambda_{\rm B}
\hbar}{4 \omega_{\alpha}} \sum_{\alpha} \frac{1}{2 \omega_{\bf k}}
= \omega_{\alpha} + \frac{\lambda_{\rm B} \hbar}{4
\omega_{\alpha}} J_0.
\end{equation}
By requiring the solution (\ref{zero sol}) to satisfy the
Wronskian condition (\ref{wron}), the solution is given by
\begin{equation}
\varphi_{\alpha} = \frac{1}{ \sqrt{2 \Omega^{(-)}_{\alpha}}} \exp
[- i \Omega^{(-)}_{\alpha} t]. \label{first sol2}
\end{equation}
The solution (\ref{first sol2}) of the MSPT method coincides with
the frequency (\ref{freq ex}) to the first order of $\lambda_{\rm
B} \hbar$. As a consequence, the MSPT solution (\ref{first sol2})
recovers the GEP (\ref{gep}) and renormalized mass (\ref{ren
mass}) to the first order $\lambda_{\rm B} \hbar$. All higher
order corrections to the MSPT solution are expected to exactly
recover the solution (\ref{sol1}) of GEP. In this sense the MSPT
method can be understood as a powerful tool to systematically
solve Eq. (\ref{cl eq}) even for the nonequilibrium case with
time-dependent coupling parameters.

We now apply the MSPT to a quantum field changing suddenly from
one parameter $m_{{\rm B}}^2$ to another $\tilde{m}_{\rm B}^{2}
\geq 0$ at $t= 0$. The mean field equation (\ref{mean eq}) still
applies with the modification
\begin{equation}
\tilde{\omega}_{\alpha}^2 = \tilde{m}_{\rm B}^{2} + {\bf k}^2,
\end{equation}
and the solution (\ref{mspt sol1}) has the form
\begin{equation}
\tilde{\varphi}_{\alpha}^{(0)} (t) = \tilde{A}_{\alpha} (\tau)
e^{i \tilde{\omega}_{\alpha} t} + \tilde{B}_{\alpha} (\tau) e^{- i
\tilde{\omega}_{\alpha} t}. \label{mspt sol3}
\end{equation}
The solution is given by
\begin{eqnarray}
\tilde{A}_{\alpha} (\tau) = \tilde{A}_{\alpha} (0) \exp \Biggl[
\frac{i \tau}{4 \tilde{\omega}_{\alpha} }\Biggl\{
\tilde{B}_{\alpha}^* (0) \tilde{B}_{\alpha} (0) +  \sum_{\beta}
\tilde{A}_{\beta}^* (0) \tilde{A}_{\beta} (0) +
\tilde{B}_{\beta}^* (0) \tilde{B}_{\beta} (0) \Biggr\}\Biggr],
\nonumber\\ \tilde{B}_{\alpha} (\tau) = \tilde{B}_{\alpha} (0)
\exp \Biggl[ \frac{- i \tau}{4 \tilde{\omega}_{\alpha} } \Biggl\{
\tilde{A}_{\alpha}^* (0) \tilde{A}_{\alpha} (0) +  \sum_{\beta}
\tilde{A}_{\beta}^* (0) \tilde{A}_{\beta} (0) +
\tilde{B}_{\beta}^* (0) \tilde{B}_{\beta} (0) \Biggr\}\Biggr].
\label{mspt sol4}
\end{eqnarray}
The initial data is determined by the continuity of
$\varphi_{\alpha}$ and $\tilde{\varphi}_{\alpha}$ at $t = 0$:
\begin{equation}
\tilde{A}_{\alpha} (0) =
\frac{1}{2\sqrt{2\omega_{\alpha}}}\Biggl(1 -
\frac{\omega_{\alpha}}{\tilde{\omega}_{\alpha}}\Biggr), \quad
\tilde{B}_{\alpha} (0) =
\frac{1}{2\sqrt{2\omega_{\alpha}}}\Biggl(1 +
\frac{\omega_{\alpha}}{\tilde{\omega}_{\alpha}}\Biggr).
\end{equation}
The zeroth order solution can then written be in terms of the
modified frequencies as
\begin{eqnarray}
\tilde{\varphi}_{\alpha}^{(0)} (\tau) &=& \tilde{A}_{\alpha} (0)
\exp {(i \tilde{\Omega}^{(+)}_{\alpha}t)} + \tilde{B}_{\alpha} (0)
\exp[-i \tilde{\Omega}^{(-)}_{\alpha}t]. \label{mspt sol5}
\end{eqnarray}
where
\begin{equation}
\tilde{\Omega}^{(\pm)}_{\alpha} = \tilde{\omega}_{\alpha} +
\frac{\lambda_{\rm B} \hbar} {4 \tilde{\omega}_{\alpha}} \Biggl[
\sum_{\beta} \frac{1}{2 \omega_{\beta}} + \frac{(m_{\rm B}^{2} -
\tilde{m}_{\rm B}^{2})}{2} \sum_{\beta} \frac{1}{2 \omega_{\beta}
\tilde{\omega}^2_{\beta}} + \frac{1}{8\omega_{\alpha}} \Biggl( 1
\pm \frac{\omega_{\alpha}}{\tilde{\omega}_{\alpha}} \Biggr)^2
\Biggr]. \label{nufreq}
\end{equation}
The shifted frequency (\ref{nufreq}) can be written as
\begin{equation}
\tilde{\Omega}^{(\pm)}_{\alpha} = \tilde{\omega}_{\alpha} +
\frac{\lambda_{\rm B} \hbar} {4 \tilde{\omega}_{\alpha}} \Biggl[
J_0 + \frac{(m_{\rm B}^{2} - \tilde{m}_{\rm B}^{2})}{2} J_{-1} +
\frac{(m_{\rm R}^2 - \tilde{m}_{\rm R}^2)^2}{2} \sum_{\beta}
\frac{1}{2 \omega^3_{\beta} \tilde{\omega}^2_{\beta}} +
\frac{1}{8\omega_{\alpha}} \Biggl( 1 \pm
\frac{\omega_{\alpha}}{\tilde{\omega}_{\alpha}} \Biggr)^2 \Biggr],
\end{equation}
where we used the relation
\begin{equation}
m^2_{\rm B} - \tilde{m}_{\rm B}^2 = m^2_{\rm R} - \tilde{m}_{\rm
R}^2
\end{equation}
and rewrote the sum
\begin{eqnarray}
\sum_{\beta} \frac{1}{2 \omega_{\beta} \tilde{\omega}^2_{\beta}}
&=& \sum_{\beta} \frac{1}{2 \omega^3_{\beta}} + \sum_{\beta}
\frac{\omega^2_{\beta} - \tilde{\omega}^2_{\beta}}{2
\omega^3_{\beta} \tilde{\omega}^2_{\beta}} \nonumber\\ &=& J_{-1}
+ (m_{\rm R}^2 - \tilde{m}_{\rm R}^2 ) \sum_{\beta} \frac{1}{2
\omega^3_{\beta} \tilde{\omega}^2_{\beta}}.
\end{eqnarray}
Finally, by fixing the coefficients of Eq. (\ref{mspt sol5}) to
match Eq. (\ref{first sol2}) and satisfy Eq. (\ref{wron}) we
obtain the lowest order solutions
\begin{eqnarray}
\tilde{\varphi}_{\alpha} (\tau) &=&
\frac{1}{\sqrt{2\Omega^{(-)}_{\alpha}}}\Biggl(\frac{
\tilde{\Omega}^{(-)}_{\alpha} -  \Omega^{(-)}_{\alpha}
}{\tilde{\Omega}^{(+)}_{\alpha} + \tilde{\Omega}^{(-)}_{\alpha}}
\Biggr) \exp [i \tilde{\Omega}^{(+)}_{\alpha}t] +
\frac{1}{\sqrt{2\Omega^{(-)}_{\alpha}}} \Biggl(\frac{
\tilde{\Omega}^{(+)}_{\alpha} +  \Omega^{(-)}_{\alpha}
}{\tilde{\Omega}^{(+)}_{\alpha} + \tilde{\Omega}^{(-)}_{\alpha}}
\Biggr) \exp[-i \tilde{\Omega}^{(-)}_{\alpha}t].
\end{eqnarray}
From the GEP calculation, Eqs. (\ref{gap}) and (\ref{gep ren}), of
Sec. II , it follows that the new renormalized frequencies can be
written as
\begin{equation}
\tilde{\Omega}^{(\pm)}_{\alpha} \simeq \Omega_{{\rm R} \alpha} +
\frac{\lambda_{\rm B} \hbar} {4 \tilde{\omega}_{\alpha}} \Biggl[
\frac{(m_{\rm B}^{2} - \tilde{m}_{\rm B}^{2})}{2} J_{-1} +
\frac{(m_{\rm R}^2 - \tilde{m}_{\rm R}^2)^2}{2} \sum_{\beta}
\frac{1}{2 \omega^3_{\beta} \tilde{\omega}^2_{\beta}} +
\frac{1}{8\omega_{\alpha}} \Biggl( 1 \pm
\frac{\omega_{\alpha}}{\tilde{\omega}_{\alpha}} \Biggr)^2 \Biggr].
\end{equation}

\section{Renormalized Solution After Phase Transition}

After the quenched phase transition at $t=0$,
the soft modes (long wavelength)
 with momenta $k < \tilde{m}_{\rm B}$ become unstable and
evolve out of equilibrium, whereas the hard modes (short wavelength)
with $k > \tilde{m}_{\rm B}$ evolve towards a new
equilibrium. The Gaussian state after phase transition
can then be determined by two sets of the auxiliary fields,
one for the stable hard modes and the other
for the unstable soft modes. The mean field equation
(\ref{mean eq}) for the hard modes then become
\begin{equation}
\ddot{\tilde{\varphi}}_{{\rm H} \alpha} + \tilde{\omega}^2_{{\rm
H}  \alpha} \tilde{\varphi}_{{\rm H}\alpha} + \frac{\lambda_{\rm
B} \hbar}{2} \Biggl(\sum_{\beta} \tilde{\varphi}_{\beta}^*
\tilde{\varphi}_{\beta} \Biggr) \tilde{\varphi}_{{\rm H}\alpha} =
0, \label{ph eq1}
\end{equation}
where
\begin{equation}
\tilde{\omega}_{{\rm H} \alpha}^2 = {\bf k}^2 - \tilde{m}_{\rm
B}^2,
\end{equation}
and for the soft modes
\begin{equation}
\ddot{\tilde{\varphi}}_{{\rm S}\alpha} - \tilde{\omega}^2_{{\rm S}
\alpha} \tilde{\varphi}_{{\rm S} \alpha} + \frac{\lambda_{\rm B}
\hbar}{2} \Biggl(\sum_{\beta} \tilde{\varphi}_{\beta}^*
\tilde{\varphi}_{\beta} \Biggr) \tilde{\varphi}_{{\rm S} \alpha} =
0, \label{ph eq2}
\end{equation}
where
\begin{equation}
\tilde{\omega}_{{\rm S} \alpha}^2 =  \tilde{m}_{\rm B}^2 - {\bf
k}^2.
\end{equation}

Once again, by introducing two independent time scales $t$
and $\tau$ in accordance with the MSPT
method, the solutions of zero order
${\cal{O}}((\lambda_B \hbar)^0)$ for the hard and
soft modes can be written as
\begin{eqnarray}
\tilde{\varphi}_{{\rm H} \alpha}^{(0)} &=& \tilde{A}_{\alpha}
(\tau) e^{i \tilde{\omega}_{{\rm H} \alpha} t} +
\tilde{B}_{\alpha} (\tau) e^{- i \tilde{\omega}_{{\rm H} \alpha}
t}, \label{ph sol1}\\ \tilde{\varphi}_{{\rm S} \alpha}^{(0)} &=&
\tilde{C}_{\alpha} (\tau) e^{ \tilde{\omega}_{{\rm S} \alpha} t} +
\tilde{D}_{\alpha} (\tau) e^{- \tilde{\omega}_{{\rm S} \alpha} t}.
\label{ph sol2}
\end{eqnarray}
Substituting these zeroth order solutions into Eqs. (\ref{ph eq1})
and (\ref{ph eq2}) and eliminating the secular terms $e^{\pm i
\tilde{\omega}_{{\rm H}\alpha} t}$ for $\tilde{\varphi}_{{\rm H}
\alpha}^{(1)}$ and $e^{\pm \tilde{\omega}_{{\rm S} \alpha} t}$ for
$\tilde{\varphi}_{{\rm S} \alpha}^{(1)}$, one finds the equations
for coefficients:
\begin{eqnarray}
\frac{\partial \tilde{A}_{\alpha}}{\partial \tau}  &=& \frac{i}{4
\tilde{\omega}_{{\rm H} \alpha}} \Biggl\{\tilde{B}_{\alpha}^*
\tilde{B}_{\alpha} + \sum_{\beta_>} \tilde{A}_{\beta}^*
\tilde{A}_{\beta} + \tilde{B}_{\beta}^* \tilde{B}_{\beta}  +
\sum_{\beta_<} \tilde{C}^*_{\beta} \tilde{D}_{\beta} +
\tilde{C}_{\beta} \tilde{D}_{\beta}^* \Biggr\}
\tilde{A}_{\alpha},\nonumber\\ \frac{\partial
\tilde{B}_{\alpha}}{\partial \tau}  &=& \frac{- i}{4 \omega_{{\rm
H} \alpha}} \Biggl\{\tilde{A}_{\alpha}^* \tilde{A}_{\alpha} +
\sum_{\beta_>} \tilde{A}_{\beta}^* \tilde{A}_{\beta} +
\tilde{B}_{\beta}^* \tilde{B}_{\beta}  + \sum_{\beta_<}
\tilde{C}^*_{\beta} \tilde{D}_{\beta} + \tilde{C}_{\beta}
\tilde{D}_{\beta}^* \Biggr\} \tilde{B}_{\alpha},\nonumber\\
\frac{\partial \tilde{C}_{\alpha}}{\partial \tau}  &=& \frac{-
1}{4 \omega_{{\rm S} \alpha}} \Biggl\{\tilde{C}_{\alpha}^*
\tilde{D}_{\alpha} + \sum_{\beta_>} \tilde{A}_{\beta}^*
\tilde{A}_{\beta} + \tilde{B}_{\beta}^* \tilde{B}_{\beta}  +
\sum_{\beta_<} \tilde{C}^*_{\beta} \tilde{D}_{\beta} +
\tilde{C}_{\beta} \tilde{D}_{\beta}^* \Biggr\}
\tilde{C}_{\alpha},\nonumber\\ \frac{\partial
\tilde{D}_{\alpha}}{\partial \tau}  &=&  \frac{1}{4 \omega_{{\rm
S} \alpha}} \Biggl\{\tilde{D}_{\alpha}^* \tilde{C}_{\alpha} +
\sum_{\beta_>} \tilde{A}_{\beta}^* \tilde{A}_{\beta} +
\tilde{B}_{\beta}^* \tilde{B}_{\beta}  + \sum_{\beta_<}
\tilde{C}^*_{\beta} \tilde{D}_{\beta} + \tilde{C}_{\beta}
\tilde{D}_{\beta}^* \Biggr\} \tilde{D}_{\alpha}. \label{ph eq3}
\end{eqnarray}
Here $\beta_>$ and $\beta_<$ denote the restricted sum (integral)
over the hard and soft modes, respectively. Another set of
equations, the complex conjugate of Eq. (\ref{ph eq3}), can be
obtained for the coefficients $\tilde{A}^*_{\alpha}$,
$\tilde{B}^*_{\alpha}$, $\tilde{C}^*_{\alpha}$ and
$\tilde{D}_{\alpha}^*$ by eliminating the secular terms for
$\tilde{\varphi}_{{\rm H} \alpha}^{*(1)}$ and
$\tilde{\varphi}_{{\rm S} \alpha}^{*(1)}$. This set of equations
together with Eq. (\ref{ph eq3}) determines the initial values of
the coefficients:
\begin{eqnarray}
\tilde{A}_{\alpha}^* (\tau) \tilde{A}_{\alpha} (\tau) &=&
 \tilde{A}_{\alpha}^* (0) \tilde{A}_{\alpha} (0), \quad
\tilde{B}_{\alpha}^* (\tau) \tilde{B}_{\alpha} (\tau) =
\tilde{B}_{\alpha}^* (0) \tilde{B}_{\alpha} (0), \nonumber\\
\tilde{C}_{\alpha}^* (\tau) \tilde{D}_{\alpha} (\tau) &=&
\tilde{C}_{\alpha}^* (0) \tilde{D}_{\alpha} (0), \quad
\tilde{C}_{\alpha} (\tau) \tilde{D}_{\alpha}^* (\tau) =
\tilde{C}_{\alpha}(0) \tilde{D}_{\alpha}^* (0). \label{ph data}
\end{eqnarray}
By using the above relations, we find the solution of
Eq. (\ref{ph eq3}) as
\begin{eqnarray}
\tilde{A}_{\alpha} (\tau) &=& \tilde{A}_{\alpha} (0) \exp
\Biggl[\frac{i \tau}{4 \omega_{{\rm H} \alpha}}
\Biggl\{\tilde{B}_{\alpha}^* \tilde{B}_{\alpha} + \sum_{\beta_>}
\tilde{A}_{\beta}^* \tilde{A}_{\beta} + \tilde{B}_{\beta}^*
\tilde{B}_{\beta}  + \sum_{\beta_<} \tilde{C}^*_{\beta}
\tilde{D}_{\beta} + \tilde{C}_{\beta} \tilde{D}_{\beta}^* \Biggr\}
\Biggr], \nonumber\\ \tilde{B}_{\alpha} (\tau) &=&
\tilde{B}_{\alpha} (0) \exp \Biggl[\frac{- i \tau}{4 \omega_{{\rm
H} \alpha}} \Biggl\{\tilde{A}_{\alpha}^* \tilde{A}_{\alpha} +
\sum_{\beta_>} \tilde{A}_{\beta}^* \tilde{A}_{\beta} +
\tilde{B}_{\beta}^* \tilde{B}_{\beta}  + \sum_{\beta_<}
\tilde{C}^*_{\beta} \tilde{D}_{\beta} + \tilde{C}_{\beta}
\tilde{D}_{\beta}^* \Biggr\} \Biggr], \nonumber\\
\tilde{C}_{\alpha} (\tau) &=& \tilde{C}_{\alpha} (0) \exp
\Biggl[\frac{- \tau}{4 \omega_{{\rm S} \alpha}}
\Biggl\{\tilde{C}_{\alpha}^* \tilde{D}_{\alpha} + \sum_{\beta_>}
\tilde{A}_{\beta}^* \tilde{A}_{\beta} + \tilde{B}_{\beta}^*
\tilde{B}_{\beta}  + \sum_{\beta_<} \tilde{C}^*_{\beta}
\tilde{D}_{\beta} + \tilde{C}_{\beta} \tilde{D}_{\beta}^* \Biggr\}
\Biggr], \nonumber\\ \tilde{D}_{\alpha} (\tau) &=&
\tilde{D}_{\alpha} (0) \exp \Biggl[\frac{\tau}{4 \omega_{{\rm S}
\alpha}} \Biggl\{\tilde{D}_{\alpha}^* \tilde{C}_{\alpha} +
\sum_{\beta_>} \tilde{A}_{\beta}^* \tilde{A}_{\beta} +
\tilde{B}_{\beta}^* \tilde{B}_{\beta}  + \sum_{\beta_<}
\tilde{C}^*_{\beta} \tilde{D}_{\beta} + \tilde{C}_{\beta}
\tilde{D}_{\beta}^* \Biggr\} \Biggr]. \label{ph sol3}
\end{eqnarray}
The initial data (\ref{ph data}) are determined by continuity of
$\varphi_{\alpha}^{(0)}$ and $\tilde{\varphi}_{\alpha}^{(0)}$:
\begin{eqnarray}
\tilde{A}_{\alpha} (0) = \frac{1}{2
\sqrt{2\omega_{\alpha}}}\Biggl(1 -
\frac{\omega_{\alpha}}{\tilde{\omega}_{{\rm H} \alpha}}\Biggr),
\quad \tilde{B}_{\alpha} (0) = \frac{1}{2
\sqrt{2\omega_{\alpha}}}\Biggl(1 +
\frac{\omega_{\alpha}}{\tilde{\omega}_{{\rm H} \alpha}}\Biggr),
\nonumber\\ \tilde{C}_{\alpha} (0) = \frac{1}{2
\sqrt{2\omega_{\alpha}}}\Biggl(1 -
i\frac{\omega_{\alpha}}{\tilde{\omega}_{{\rm S} \alpha}}\Biggr),
\quad \tilde{D}_{\alpha} (0) = \frac{1}{2
\sqrt{2\omega_{\alpha}}}\Biggl(1 +
i\frac{\omega_{\alpha}}{\tilde{\omega}_{{\rm S} \alpha}}\Biggr).
\label{ph init3}
\end{eqnarray}

It is worth noting that the integral for the soft modes, the last
two terms of the summation in the exponent of Eq. (\ref{ph sol3}),
is a restricted one with its upper limit being $\tilde{m}_{\rm
B}$. On the other hand, the integral for the soft modes, the first
two terms of the summation in the exponent of Eq. (\ref{ph sol3}),
has $\tilde{m}_{\rm B}$ as its lower limit of integration. For $k
\gg \tilde{m}_{\rm B}$, one can write $\tilde{\omega}_{{\rm H}
\beta} \sim \omega_{\beta}$, whence the ultraviolet (quadratic)
divergence mainly comes from $\sum_{\beta} \tilde{B}_{\beta}^*
\tilde{B}_{\beta}$ for the hard modes. In terms of new frequencies
$\tilde{\Omega}_{{\rm H} \alpha}^{(\pm)}$ for the hard modes and
$\tilde{\Omega}_{{\rm S} \alpha}^{(\pm)}$ for the soft modes, the
zeroth order solution can be written as
\begin{eqnarray}
\tilde{\varphi}_{{\rm H} \alpha}^{(0)} &=& \tilde{A}_{\alpha}(0)
e^{i\tilde{\Omega}_{{\rm H} \alpha}^{(+)} t} +
\tilde{B}_{\alpha}(0) e^{- i \tilde{\Omega}_{{\rm H}
\alpha}^{(-)}t}, \nonumber\\ \tilde{\varphi}_{{\rm S}
\alpha}^{(0)} &=& \tilde{C}_{\alpha}(0) e^{\tilde{\Omega}_{{\rm S}
\alpha}^{(+)}t} + \tilde{D}_{\alpha}(0) e^{\tilde{\Omega}_{{\rm S}
\alpha}^{(-)}t}
\end{eqnarray}
where
\begin{eqnarray}
\tilde{\Omega}_{{\rm H} \alpha}^{(\pm)} &=& \tilde{\omega}_{{\rm
H} \alpha} + \frac{\lambda_{\rm B} \hbar}{4 \tilde{\omega}_{{\rm
H} \alpha}} \Biggl[\sum_{\beta} \frac{1}{2\omega_{\beta}} +
\frac{(m_{\rm B}^{2} + \tilde{m}_{\rm B}^{2})}{2} \Biggl\{
\sum_{\beta_>} \frac{1}{2 \omega_{\beta} \tilde{\omega}^2_{{\rm H}
\beta}} - \sum_{\beta_<} \frac{1}{2 \omega_{\beta}
\tilde{\omega}^2_{{\rm S} \beta}} \Biggr\}
 + \frac{1}{8 \omega_{\alpha}} \Biggl(1 \pm
\frac{\omega_{\alpha}}{\tilde{\omega}_{{\rm H} \alpha}} \Biggr)^2
\Biggr], \nonumber\\ \tilde{\Omega}_{{\rm S} \alpha}^{(\pm)} &=&
\tilde{\omega}_{{\rm S} \alpha} - \frac{\lambda_{\rm B} \hbar}{4
\tilde{\omega}_{{\rm S} \alpha}} \Biggl[\sum_{\beta}
\frac{1}{2\omega_{\beta}} + \frac{(m_{\rm B}^{2} + \tilde{m}_{\rm
B}^{2})}{2} \Biggl\{ \sum_{\beta_>} \frac{1}{2\omega_{\beta}
\tilde{\omega}^2_{{\rm H} \beta}} - \sum_{\beta_<}
\frac{1}{2\omega_{\beta} \tilde{\omega}^2_{{\rm S} \beta}}
\Biggr\} + \frac{1}{8 \omega_{\alpha}} \Biggl(1 \pm i
\frac{\omega_{\alpha}}{\tilde{\omega}_{{\rm S} \alpha}} \Biggr)^2
 \Biggr].
\label{ren freq}
\end{eqnarray}
Here $\sum_{\beta_<}$ denotes the restricted integral over soft
modes, having a finite upper momentum cut-off given by
$\tilde{m}_{\rm B}$ and, therefore, making a finite contribution
to the renormalized frequencies. However, the quadratic
ultraviolet divergent term has to be appropriately regularized to
yield the renormalized frequencies. Contrary to a naive belief
that the frequency of the soft modes may be finite, the soft
frequency also has the same ultraviolet divergence as the hard
modes and renormalization is required to remove it. This is a
consequence of the nonlinear coupling between the soft and the
hard modes. As the mass counter term $\delta m^2$ has the same
sign, $\tilde{m}_{\rm B}^2 = \tilde{m}_{\rm R}^2 - \delta m^2$,
even after the phase transition, the sum of bare squared masses
after the phase transition equals to that of renormalized squared
masses:
\begin{equation}
m_{\rm B}^2 + \tilde{m}_{\rm B}^2 = m_{\rm R}^2 + \tilde{m}_{\rm
R}^2.
\end{equation}
Using the identity
\begin{eqnarray}
\sum_{\beta_>} \frac{1}{2 \omega_{\beta} \tilde{\omega}_{{\rm
H}\beta}^2} &=& \sum_{\beta_>} \frac{1}{2 \omega_{\beta}^3} +
(m^2_{\rm R} + \tilde{m}_{\rm R}^2 ) \sum_{\beta_>} \frac{1}{2
\omega_{\beta}^3 \tilde{\omega}_{{\rm H} \beta}^2}, \nonumber\\
\sum_{\beta<} \frac{1}{2 \omega_{\beta} \tilde{\omega}_{{\rm
S}\beta}^2} &=& - \sum_{\beta_<} \frac{1}{2 \omega_{\beta}^3} +
(m^2_{\rm R} + \tilde{m}_{\rm R}^2 ) \sum_{\beta_<} \frac{1}{2
\omega_{\beta}^3 \tilde{\omega}_{{\rm S} \beta}^2},
\end{eqnarray}
we write the shifted frequency as
\begin{eqnarray}
\tilde{\Omega}_{{\rm H} \alpha}^{(\pm)} &=& \tilde{\omega}_{{\rm
H} \alpha} + \frac{\lambda_{\rm B} \hbar}{4 \tilde{\omega}_{{\rm
H} \alpha}} \Biggl[J_0 + \frac{(m^2_{\rm R} + \tilde{m}_{\rm
R}^2)}{2} J_{-1} + \frac{(m^2_{\rm R} + \tilde{m}_{\rm R}^2)^2}{2}
\Biggl\{ \sum_{\beta_>} \frac{1}{2\omega^3_{\beta}
\tilde{\omega}^2_{{\rm H} \beta}} - \sum_{\beta_<}
\frac{1}{2\omega^3_{\beta} \tilde{\omega}^2_{{\rm S} \beta}}
\Biggr\} + \frac{1}{8 \omega_{\alpha}} \Biggl(1 \pm
\frac{\omega_{\alpha}}{\tilde{\omega}_{{\rm H} \alpha}} \Biggr)^2
\Biggr] \nonumber\\ \tilde{\Omega}_{{\rm S} \alpha}^{(\pm)} &=&
\tilde{\omega}_{{\rm S} \alpha} - \frac{\lambda_{\rm B} \hbar}{4
\tilde{\omega}_{{\rm S} \alpha}} \Biggl[
 J_0 + \frac{(m^2_{\rm R} + \tilde{m}_{\rm
R}^2)}{2} J_{-1} + \frac{(m^2_{\rm R} + \tilde{m}_{\rm R}^2)^2}{2}
\Biggl\{ \sum_{\beta_>} \frac{1}{2\omega^3_{\beta}
\tilde{\omega}^2_{{\rm H} \beta}} - \sum_{\beta_<}
\frac{1}{2\omega^3_{\beta} \tilde{\omega}^2_{{\rm S} \beta}}
\Biggr\} + \frac{1}{8 \omega_{\alpha}} \Biggl(1 \pm i
\frac{\omega_{\alpha}}{\tilde{\omega}_{{\rm S} \alpha}} \Biggr)^2
\Biggr]. \label{ren freq3}
\end{eqnarray}
The first two terms of the MSPT frequency (\ref{ren freq3})
\begin{equation}
\tilde{\Omega}_{\alpha}^{(\pm)} \approx \tilde{\omega}_{\alpha}
\pm \frac{\lambda_{\rm B} \hbar}{4 \tilde{\omega}_{\alpha}} J_0
\end{equation}
gives, to the first order of $\lambda_{\rm B} \hbar$, the
renormalized frequency
\begin{equation}
\tilde{\Omega}_{{\rm R} \alpha}^2 \approx
\tilde{\omega}_{\alpha}^2 \pm \frac{\lambda_{\rm B} \hbar }{2}
J_{0}.
\end{equation}
Finally, the lowest order solutions which match the solutions
(\ref{first sol2}) before phase transition and satisfy Eq.
(\ref{wron}), are given by
\begin{eqnarray}
\tilde{\varphi}_{{\rm H} \alpha} &=& \frac{1}{
\sqrt{2\Omega^{(-)}_{\alpha}}}\Biggl(\frac{\tilde{\Omega}^{(-)}_{{\rm
H}\alpha} - \Omega^{(-)}_{\alpha}}{\tilde{\Omega}^{(+)}_{{\rm
H}\alpha} + \tilde{\Omega}^{(-)}_{{\rm H} \alpha}} \Biggr) e^{i
\tilde{\Omega}_{{\rm H} \alpha}^{(+)} t} + \frac{1}{
\sqrt{2\Omega^{(-)}_{\alpha}}}
\Biggl(\frac{\tilde{\Omega}^{(+)}_{{\rm H}\alpha} -
\Omega^{(-)}_{\alpha}}{\tilde{\Omega}^{(+)}_{{\rm H}\alpha} +
\tilde{\Omega}^{(-)}_{{\rm H} \alpha}} \Biggr) e^{- i
\tilde{\Omega}_{{\rm H} \alpha}^{(-)}t}, \nonumber\\
\tilde{\varphi}_{{\rm S} \alpha} &=&
\frac{1}{\sqrt{2\Omega^{(-)}_{\alpha}}}
\Biggl(\frac{\tilde{\Omega}^{(-)}_{{\rm S}\alpha} - i
\Omega^{(-)}_{\alpha}}{\tilde{\Omega}^{(+)}_{{\rm S}\alpha} +
\tilde{\Omega}^{(-)}_{{\rm S} \alpha}} \Biggr)
e^{\tilde{\Omega}_{{\rm S} \alpha}^{(+)}t} + \frac{1}{
\sqrt{2\Omega^{(-)}_{\alpha}}}\Biggl(\frac{\tilde{\Omega}^{(+)}_{{\rm
S}\alpha} + i \Omega^{(-)}_{\alpha}}{\tilde{\Omega}^{(+)}_{{\rm
S}\alpha} + \tilde{\Omega}^{(-)}_{{\rm S} \alpha}} \Biggr)
e^{\tilde{\Omega}_{{\rm S} \alpha}^{(-)}t}
\end{eqnarray}
It is shown that the MSPT method, when applied to the mean field
equation (\ref{mean eq}), yields the correct renormalized
frequency even after the phase transition.

\section{Conclusion}

In summary, we have studied certain aspects of the renormalization
of quantum fields evolving out of equilibrium using the recently
developed Liouville-von Neumann method \cite{kim-lee2}. The
multiple scale perturbation theory (MSPT) provides a powerful and
useful tool in finding an analytic approximate solution to
nonlinear field equations. The mean-field equations for the
auxiliary variables for the Gaussian state, which are equivalent
to the vacuum expectation of the Hartree-Fock equation, are
analyzed using the MSPT. As a simple quantum mechanical model, two
coupled quartic oscillators are studied. In particular, the model
with one unstable quartic oscillator, mimicking a soft mode of a
nonequilibrium field, sheds some light on the effect of nonlinear
mode coupling between the soft and the hard modes in the
nonequilibrium evolution of phase transitions \cite{ksk}. The MSPT
method is employed to the $\Phi^4$-theory before the phase
transition and its lowest order solutions to the mean-field
equations involve the correct renormalized frequency consistent
with the Gaussian effective potential \cite{stevenson}. Also the
MSPT method turns out to be a powerful and practical tool for the
renormalization of nonequilibrium quantum fields such as they
appear in phase transitions.

Although we have used Eq. (\ref{ren mass}) in order to express the
modified frequencies in terms of the finite renormalized
parameters, there is a caveat which is necessary to keep in mind.
The renormalized mass is obtained from the quantum fluctuations of
the Gaussian state around $\phi_c = 0$ for the static and stable
$\Phi^4$-theory. When the symmetry is spontaneously broken,
$\phi_c = 0$ is no longer the true vacuum state. However, the
justification of Eq. (\ref{ren mass}) in obtaining renormalized
parameters, is solely dependent on the same Gaussian approximation
whose quantum state is the time-dependent Gaussian state around
$\phi_c = 0$ throughout the phase transition. This effectively
amounts to studying the nonequilibrium field dynamics during the
rolling down of the field and before the fluctuations start
probing the true vacuum state. As far as the soft modes are
concerned, they just see an inverted potential. Therefore, the
lowest order solution of the MSPT method exponentially grows
indefinitely and begins to oscillate slowly. The imaginary part of
the solution, leading to the oscillation, determines the width of
the Gaussian state at later times, roughly covering the true
vacuum state. The exponential growth of the lowest order solution
is terminated by the first order correction of the MSPT method
when the field begins to sample around the true vacuum state
\cite{ksk}.

In this paper we confined our attention to the Gaussian
approximation using the LvN formalism (which is equivalent to the
Hartree-Fock or mean-field method) to address issues related to
renormalization of a spontaneously broken $\Phi^4$-theory. We now
point out some of the limitations of the Hartree-Fock method since
an understanding of those limitations is crucial in developing
techniques for more accurately describing nonequilibrium dynamics
of phase transitions. As has been recognized in the literature,
the Hartree-Fock method (sometimes called the {\it collisionless}
Hartree-Fock method) neglects the effects of scattering. It also
cannot account for late time thermalization of the system since it
takes into account the effect of interactions only through a mean
field. It is based on the assumption that scattering effects do
not play a very crucial role in the early stages of phase
transition dynamics. However, recent studies \cite{berges}
indicate that scattering effects may be crucial to understanding
the dynamical evolution of the system far from equilibrium.
Similarly, in the Liouville-von Neumann method, the quartic part
$\hat{H}_{\rm P}$, which is a perturbation to the quadratic
(Gaussian) part $\hat{H}_{\rm G}$ and is negligible before the
phase transition, grows comparable to $\hat{H}_{\rm G}$ and
significantly contributes to the Gaussian state
\cite{kim-khanna2}. These terms may correspond to the
next-to-leading order contribution to the effective action of Ref.
\cite{berges}. The study of non-equilibrium phase transition
dynamics beyond the Gaussian approximation is under investigation.

\acknowledgements

This work was supported in part by the Natural Sciences and
Engineering Research Council of Canada. The work of S.P.K. was
also supported in part by the Korea Research Foundation under
Grant No. 1998-001-D00364.

\end{document}